\documentclass[aps,prd,eqsecnum,nofootinbib,twocolumn]{revtex4-2}
\usepackage{amsmath,amssymb,bm}
\usepackage[dvips]{graphicx}
\usepackage{color}
\newcommand{\be}{\begin{equation}}
\newcommand{\ee}{\end{equation}}
\newcommand{\bea}{\begin{eqnarray}}
\newcommand{\eea}{\end{eqnarray}}

\newcommand{\eq}[1]{Eq.~(\ref{eq:#1})}
\newcommand{\sect}[1]{Sec.~\ref{sec:#1}}
\newcommand{\appen}[1]{Appendix~\ref{sec:#1}}

\newcommand{\del}{\partial}

\newcommand{\bra}{\langle}
\newcommand{\ket}{\rangle}
\newcommand{\calO}{{\cal O}}

\newcommand{\eg}{{\it e.g.}}
\newcommand{\ie}{{\it i.e.}}

\bmdefine{\bmq}{{\bm{q}}}
\bmdefine{\bmk}{{\bm{k}}}
\bmdefine{\bmx}{{\bm{x}}}
\bmdefine{\bmy}{{\bm{y}}}
\bmdefine{\bmr}{{\bm{r}}}
\bmdefine{\bmnabla}{{\bm{\nabla}}}
\bmdefine{\bmA}{ \bm{A} }
\bmdefine{\bmD}{ \bm{D} }

\bmdefine{\bmPhi}{ \bm{\Phi} }
\bmdefine{\bmPsi}{ \bm{\Psi} }
\bmdefine{\bmcalO}{ \bm{\mathcal{O}} }

\newcommand{\calM}{{\cal M}}

\newcommand{\tilx}{\tilde{x}}

\newcommand{\vecx}{\vec{x}}

\newcommand{\nq}{\mathfrak{q}}

\newcommand{\nw}{\mathfrak{w}}

\newcommand{\Deltap}{\Delta_+}
\newcommand{\Deltam}{\Delta_-}
\newcommand{\Deltapm}{\Delta_\pm}

\bmdefine{\bmg}{{\bm{g}}}
\bmdefine{\bmR}{{\bm{R}}}

\newcommand{\GR}{G^R}

\newcommand{\ro}{r_+}
\newcommand{\ri}{r_-}
\newcommand{\TL}{T_L}
\newcommand{\TR}{T_R}

\newcommand{\nlp}{n_L^p} %n_p
\newcommand{\nlz}{n_L^z} %n_z
\newcommand{\nrp}{n_R^p} %n_p'
\newcommand{\nrz}{n_R^z} %n_z'
\newcommand{\mlz}{m_L^z} %m_z
\newcommand{\mrz}{m_R^z} %m_z'

\newcommand{\cn}{n}
\newcommand{\half}{\tfrac{1}{2}}

\newcommand{\hatt}{\hat{t}}
\newcommand{\hx}{\hat{x}}
\newcommand{\hr}{\hat{r}}
\newcommand{\hT}{\hat{T}}
\newcommand{\homega}{\hat{\omega}}
\newcommand{\hq}{\hat{q}}

\newcommand{\bwt}{\begin{widetext}}
\newcommand{\ewt}{\end{widetext}}

\begin{document}
%\begin{CJK*}{GB}{} % Use default fonts from CJK (see below)

%%%   Title Page   %%%

\title{Pole-skipping and zero temperature}
\author{Makoto Natsuume}%(夏梅誠)]
\email{makoto.natsuume@kek.jp}
\altaffiliation[Also at]{
Department of Particle and Nuclear Physics, 
SOKENDAI (The Graduate University for Advanced Studies), 1-1 Oho, 
Tsukuba, Ibaraki, 305-0801, Japan;
 Department of Physics Engineering, Mie University, 
 Tsu, 514-8507, Japan.}
\affiliation{KEK Theory Center, Institute of Particle and Nuclear Studies, 
High Energy Accelerator Research Organization,
Tsukuba, Ibaraki, 305-0801, Japan}
\author{Takashi Okamura}%(岡村隆)
\email{tokamura@kwansei.ac.jp}
\affiliation{Department of Physics, Kwansei Gakuin University,
Sanda, Hyogo, 669-1337, Japan}
\date{\today}
\begin{abstract}
We study the pole-skipping phenomenon of the scalar retarded Green's function in the rotating BTZ black hole background. In the static case, the pole-skipping points are typically located at negative imaginary Matsubara frequencies $\omega = -(2\pi T)ni$ with appropriate values of complex wave number $q$. But, in a $(1+1)$-dimensional CFT, one can introduce temperatures for left-moving and right-moving sectors independently. As a result, the pole-skipping points $\omega$ depend both on left and right temperatures in the rotating background. In the extreme limit, the pole-skipping does not occur in general. But in a special case, the pole-skipping does occur even in the extreme limit, and the pole-skipping points are given by right Matsubara frequencies. 
\end{abstract}
%
% KEK-TH-2277

\maketitle
%\end{CJK*}

%%%%%%%%%
\section{Introduction and Summary}%\label{sec:}
%%%%%%%%%
The AdS/CFT duality or holography \cite{Maldacena:1997re,Witten:1998qj,Witten:1998zw,Gubser:1998bc} is a useful tool to study strongly-coupled systems (see, \eg, Refs.~\cite{CasalderreySolana:2011us,Natsuume:2014sfa,Ammon:2015wua,Zaanen:2015oix,Hartnoll:2016apf}). Recently, it is found that many finite-temperature Green's functions are not unique at ``special points" or ``pole-skipping points" in complex momentum space $(\omega, q)$, where $\omega$ is frequency and $q$ is wave number \cite{Grozdanov:2019uhi,Blake:2019otz,Natsuume:2019xcy}. Such a phenomenon is collectively known as ``pole-skipping."

The pole-skipping was originally discussed in the context of holographic chaos \cite{Shenker:2013pqa,Roberts:2014isa,Roberts:2014ifa,Shenker:2014cwa,Maldacena:2015waa}. In this case, the pole-skipping was observed in the energy density Green's function which is extracted from the bulk gravitational sound mode (see, \eg, Refs.~\cite{Grozdanov:2017ajz,Blake:2018leo,Grozdanov:2018kkt,Natsuume:2019sfp}). However, it turns out that various other bulk fields also show this behavior \cite{Grozdanov:2019uhi,Blake:2019otz,Natsuume:2019xcy}. Since then, various aspects of the pole-skipping have been investigated (see, \eg, Refs.~\cite{Natsuume:2019vcv,Wu:2019esr,Balm:2019dxk,Ceplak:2019ymw,Ahn:2019rnq,Ahn:2020bks,Abbasi:2020ykq,Jansen:2020hfd,Ahn:2020baf}).
%Various Green's functions exhibit this phenomenon. In addition to the gravitational sound mode (energy density correlators) which was originally discussed in holographic chaos, the bulk scalar field, the bulk Maxwell field (the current and charge correlators from the boundary point of view), the gravitational shear mode (momentum correlators) and the tensor mode show this behavior. 

There is a universality for the pole-skipping points $\omega$. In all examples, pole-skipping points are located at Matsubara frequencies. Typically, they start from $\nw := \omega/(2\pi T) =-i $ and continue $\nw_n = -in $ for a positive integer $n$%
\footnote{For spinors, the pole-skipping points are given by fermionic Matsubara frequencies $\omega_n=-i(2\pi T)(n+1/2)$ \cite{Ceplak:2019ymw}.}. On the other hand, the value of $q_n$ depends on the system. 
For the sound mode, pole-skipping points start from $\nw_{-1} = +i$. It is argued that the $\nw_{-1} = +i$ pole-skipping point is related to a chaotic behavior (see \sect{previous} for more details). 

The Green's function is not unique because the residue of a pole vanishes at a pole-skipping point. Generically, one would write a Green's function as
\begin{align}
\GR (\omega,q) = \frac{b(\omega,q)}{a(\omega,q)}~.
\label{eq:2pt}
\end{align}
At pole-skipping points, $a(\omega_n, q_n)=b(\omega_n, q_n)=0$. 
%Near the special point, the Green's function is then expanded as
%\begin{subequations}
%\label{eq:}
%\begin{align}
%
%\GR &= \frac{ \delta\omega (\del_\omega b)_n + \delta q (\del_q b)_n +\cdots }{\delta\omega (\del_\omega a)_n + \delta q (\del_q a)_n +\cdots } 
%\\
%&= \frac{ (\del_\omega b)_n + \frac{\delta q}{\delta\omega} (\del_q b)_n +\cdots }{ (\del_\omega a)_n + \frac{\delta q}{\delta\omega} (\del_q a)_n +\cdots }~.
%\label{eq:}
%
%\end{align}
%\end{subequations}
For example, consider a scalar field in the BTZ black hole background \cite{Banados:1992wn,Banados:1992gq}. The first pole-skipping point is located at $(\nw_1, \nq_1)=(-i,i\nu)$, where $\nu=\sqrt{1+m^2}$. Near the pole-skipping point, the Green's function behaves as
\begin{align}
\GR \propto \frac{\delta \omega+\delta q}{\delta \omega-\delta q}~,
\label{eq:pole_skip_example}
\end{align}
and the Green's function is not uniquely determined. Rather, it depends on the slope $\delta q/\delta\omega$.

In this paper, we consider the rotating BTZ black hole background which is dual to a $(1+1)$-dimensional conformal field theory. We study the pole-skipping of the minimally-coupled scalar field in the background. The Green's function is known in exact analytic form \cite{Birmingham:2001pj,Son:2002sd}. The rotating black hole gives interesting lessons about the pole-skipping: 
\begin{enumerate}
\item Unlike the static case, the pole-skipping points $\omega$ do not take the simple form.
\item In the extreme limit, the pole-skipping does not occur in general. 
\item However, there is an exceptional case, and the pole-skipping does occur even in the extreme limit. 
\end{enumerate}

In a $(1+1)$-dimensional CFT, one can introduce temperatures for left-moving and right-moving sectors independently, $\TL$ and $\TR$. The pole-skipping points $\omega$ depend both on $\TL$ and $\TR$ (Point~1). In addition, $\omega$ also depends on the conformal dimension $\Deltap=1+\nu$ of the dual boundary theory operator $\calO$%
%v2
\footnote{
One often employs the so-called comoving coordinate in the rotating background [see \eq{comoving}]. In this coordinate system, pole-skipping points $\omega$ take the simple form and is given by Matsubara frequencies. However, the comoving coordinate system represents the rotating boundary metric whereas the original coordinate system represents the nonrotating boundary metric. So, these coordinate systems have different boundary interpretations (\sect{rotating_BTZ} and \ref{sec:pole_skip}). 
}.
%However, in the so-called comoving coordinate [see \eq{comoving}], $\omega$ takes the simple form and is given by Matsubara frequencies.
In holographic chaos, it is known that Lyapunov exponents depend both on $\TL$ and $\TR$ though the studies of the out-of-time-ordered correlators (OTOC) and the pole-skipping in the rotating BTZ background \cite{Poojary:2018esz,Jahnke:2019gxr,Mezei:2019dfv,Banerjee:2019vff,Liu:2020yaf}.

For Point~2, the standard pole-skipping analysis (reviewed in \sect{pole-skipping}) utilizes a power-series expansion around the horizon $r=\ro$. Namely, one constructs perturbative solutions via Frobenius' method:
\begin{align}
\phi (r) =\sum_{n=0}^\infty \phi_n(r-\ro)^{n+\lambda}~.
%\label{eq:}
%
\end{align}
But this method does not work for extreme black holes. An extreme black hole has a degenerate horizon, and the perturbation equation has an irregular singularity there. In order for Frobenius' method to work, one has to expand the equation around a regular singularity or a regular point%
\footnote{More precisely, Frobenius' method may give a solution even around the degenerate horizon but does not give 2 solutions. Also, the solution typically has zero radius of convergence. }.
%Also, the closest singularity from the outer horizon $r=\ro$ is the inner horizon $r=\ri$, and they coincide for a degenerate horizon, so the power-series solution has zero radius of convergence. (?)}. 
Thus, the standard pole-skipping analysis does not work in the extreme limit. 

Because the pole-skipping points are given by $\omega_n=-(2\pi T)ni$, one would conclude no nontrivial pole-skipping (except $\omega=q=0$) by taking the $T\to0$ limit naively. This is not correct from two reasons. First, the pole-skipping analysis and the extreme limit do not commute (\sect{stationary_pole_skip}). Indeed, in this example, even $\omega=q=0$ is not a pole-skipping point. Second, in a $(1+1)$-dimensional CFT, one has both $\TL$ and $\TR$. Thus, even in the extreme limit $\TL\to0$, the temperature dependence remains in the Green's function, and there is a possibility of the extreme pole-skipping. 

There is another issue about the extreme pole-skipping. In general, Green's functions have lines of poles at finite temperatures. These poles contribute to the pole-skipping. However, in the zero-temperature limit, these poles are replaced by branch cuts. Thus, the pole-skipping is not clear in the extreme limit. 

%v2
For the BTZ black hole, analytic Green's function %is known
can be derived both for the nonextreme case and for the extreme case, so one can study the pole-skipping in the extreme limit, and there is no pole-skipping in general (Point~2). 
%In the extreme limit, there is no pole-skipping in general. 
However, there is an exceptional case, and the pole-skipping does occur even in the extreme limit (Point~3). This occurs when $\nu=1$, and the pole-skipping points $\omega$ are given by {\it right Matsubara frequencies}.

The reason of these phenomena is simple. In this example, the pole-skipping occurs when left poles (zeros) coincide with right zeros (poles). The pole-skipping never occurs in each sector alone. 
%\margin{?}
Thus, the pole-skipping points in general depend on $\TL$ and $\TR$. 
In the extreme limit $\TL\to0$, the left contribution gives only a power-law behavior $(\omega-q)^\nu$ in the Green's function (with possible log terms when there is matter conformal anomaly), and there is no nontrivial poles nor zeros. Thus, there is no pole-skipping in general. Note that the power-law is the expected behavior for a CFT, but the full Green's function still has nontrivial poles and zeros from the right sector. 
The reason of the exceptional case is also simple. The power-law $(\omega-q)^\nu$ becomes a simple zero when $\nu=1$, and the pole-skipping occurs from the left zero and right poles. 

To summarize, the pole-skipping seems 
%v2
to be a finite-temperature phenomenon, and it does not occur at zero temperature in general. But, in the extreme limit, one still has the nonvanishing right temperature $\TR$, so the pole-skipping is possible. 
%We imagine that this is true in general. 

%The plan of this paper is as follows...

%%%%%%%%%
\section{Rotating BTZ black hole}\label{sec:rotating_BTZ}
%%%%%%%%%

The rotating BTZ black hole is given by
\begin{subequations}
\label{eq:sch}
\begin{align}
ds^2 &= -Fdt^2+\frac{dr^2}{F}+r^2 \left(dx-\frac{\ro\ri}{r^2}dt \right)^2~, \\
F&=\frac{(r^2-\ro^2)(r^2-\ri^2)}{r^2}~.
%\label{eq:}
%
\end{align}
\end{subequations}
For simplicity, we set the AdS radius $L=1$. 
We call this coordinates the ``BTZ coordinates."
For the BTZ black hole, $x$ is actually an angular coordinate and is compact $x\approx x+2\pi$, but we regard $x$ as a noncompact coordinate. 
The temperature $T$, mass $M$, angular momentum $J$, and angular momentum potential $\Omega$ are given by 
\begin{subequations}
%\label{eq:}
\begin{align}
2\pi T&=\frac{\ro^2-\ri^2}{\ro}~, \quad
M=\ro^2+\ri^2~,\\
J&=2\ro\ri~, \quad
\Omega = \frac{\ri}{\ro}~,
%\\S= \frac{2\pi\ro}{4G} = 4\pi\ro = 4\pi^2(\TL+\TR)~,
%\label{eq:}
%
\end{align}
\end{subequations}
(We set $8G=1$ following the standard convention.)

A two-dimensional CFT has two independent modes, left-movers and right-movers, and one can introduce temperatures for each sectors:
\begin{subequations}
%\label{eq:}
\begin{align}
2\pi \TL &= \frac{2\pi T}{1+\Omega} =\ro-\ri~, \\
2\pi \TR &= \frac{2\pi T}{1-\Omega} =\ro+\ri~,\\
\frac{2}{T} &= \frac{1}{\TL}+\frac{1}{\TR}~, \\
\Omega &= \frac{\TR-\TL}{\TR+\TL}~.
%\label{eq:}
%
\end{align}
\end{subequations}
Following the standard convention, we choose $T_L\to0$ in the extreme limit. 

In the rotating BTZ black hole, the comoving coordinate is often used: it is given by 
%v2
the Galilean boost on the boundary coordinate:
\begin{align}
x' = x-\frac{\ri}{\ro} t~,
\label{eq:comoving}
\end{align}
and the metric becomes
\begin{align}
ds^2 &= -Fdt^2+\frac{dr^2}{F}+r^2 \left( \frac{\ri}{\ro}\frac{r^2-\ro^2}{r^2}dt + dx' \right)^2~.
%\label{eq:}
%
\end{align}
This coordinate comoves with the outer horizon so that the metric is diagonal at the horizon. 
We consider the perturbation of the form $e^{-i\omega t+iqx}$ in the BTZ coordinates. In the comoving coordinate,
\begin{align}
\omega' =\omega -\frac{\ri}{\ro}q~, \quad q'=q~.
%\label{eq:}
%
\end{align}
Since the comoving metric asymptotically behaves as
\begin{align}
ds^2 &\stackrel{r\to\infty}{\longrightarrow} r^2[ -dt^2+(\Omega dt+dx')^2 ]~,
%\label{eq:}
%
\end{align}
it represents the rotating boundary metric. On the other hand, in the BTZ coordinates \eqref{eq:sch}, the metric asymptotically behaves as
\begin{align}
ds^2 &\stackrel{r\to\infty}{\longrightarrow} r^2[ -dt^2+dx^2 ]~,
%\label{eq:}
%
\end{align}
so it represents the nonrotating boundary metric. 

In the pole-skipping analysis, it is convenient to use the incoming Eddington-Finkelstein (EF) coordinates. In the rotating case, 
\begin{subequations}
%\label{eq:}
\begin{align}
dr_* &:= \frac{dr}{F}~, \\%\quad
dv &:= dt +dr_*~, \quad
d\tilx := dx+ \frac{\ro\ri}{r^2F} dr~,
\label{eq:tortoise}
\end{align}
\end{subequations}
and the metric becomes
\begin{subequations}
%\label{eq:}
\begin{align}
ds^2 =& -Fdv^2 + 2dvdr + r^2 \left( d\tilx - \frac{\ro\ri}{r^2}dv \right)^2 \\
=& -\{r^2-(\ro^2+\ri^2)\}dv^2 + 2dv \left(dr - \frac{\ri}{\ro} d\tilx \right) 
\nonumber \\
&+ r^2d\tilx^2~.
\label{eq:EF}
\end{align}
\end{subequations}

Let us focus on the asymptotic behavior. We consider the asymptotically AdS background
%The metric behaves as
\begin{subequations}
%\label{eq:}
\begin{align}
ds^2 &\sim -r^2 dt^2 + \frac{dr^2}{r^2} +r^2 dx^2 \\
&=\frac{1}{u^2}(-dt^2+du^2+dx^2)~,
%\label{eq:}
%
\end{align}
\end{subequations}
where $u:=1/r$. 
We consider the minimally-coupled scalar field $\phi$ in the background 
and consider the perturbation of the form $\phi(u)e^{-i\omega t+iqx}$. 
%From the scalar field equation $(\nabla^2-m^2)\phi=0$, 
The perturbation behaves as 
\begin{subequations}
%\label{eq:}
\begin{align}
\phi &\sim A u^{\Deltam} + B u^{\Deltap}~,
\quad (u\to0)~,\\
\Delta_\pm &=1\pm \nu~, \\
\nu &:= \sqrt{1+m^2}~.
%\label{eq:}
%
\end{align}
\end{subequations}
According to the standard AdS/CFT dictionary, the slow falloff $A$ represents the source of the boundary operator $\calO$, and the fast falloff $B$ represents the response to $\calO$:
\begin{align}
\bra\calO\ket = 2\nu B~.
%\label{eq:}
%
\end{align}
Then, the retarded Green's function $\GR$ is given by
\begin{align}
\GR = -\frac{\delta\bra\calO\ket}{\delta A} = -(2\nu)\frac{B}{A}~.
%\label{eq:}
%
\end{align}
This is for the standard quantization \cite{Klebanov:1999tb}, and $\calO$ has conformal dimension $\Deltap$. In the alternative quantization, the role of $A$ and $B$ is exchanged. We assume $\calO$ has a positive conformal dimension. 

When $\nu=\mathbb{Z}$, the asymptotic behavior and the AdS/CFT dictionary are modified. See \appen{integer_Green}.

%%%%%%%%%
\section{Scalar Green's function and pole-skipping}\label{sec:stationary_pole_skip}
%%%%%%%%%

%We consider the minimally-coupled scalar field $\phi$ and consider the perturbation of the form $e^{-i\omega t+iqx}$. The scalar field is dual to the boundary operator $\calO$ and we consider its Green's function. The operator has conformal dimension $\Deltap=1+\nu$ where $\nu:=\sqrt{1+m^2}$. 
In this section, we assume $\nu\neq\mathbb{Z}$. The  $\nu=\mathbb{Z}$ case is discussed separately in next section.

%In \appen{stationary_Green}, the scalar field equation $(\nabla^2-m^2)\phi=0$ is solved in the rotating BTZ background, and the retarded Green's function is obtained:
%v2
The scalar Green's function can be obtained by solving the scalar field equation $(\nabla^2-m^2)\phi=0$ in the rotating BTZ background. This was originally derived in Refs.~\cite{Birmingham:2001pj,Son:2002sd} and was reproduced in many literature, but we rederive it for completeness in \appen{stationary_Green}. The Green's function is given by
\begin{align}
\GR =& -2\nu(4\pi^2\TL\TR)^\nu 
\nonumber \\
&\times
\frac{ \Gamma(-\nu) }{ \Gamma(\nu) }
\frac{\Gamma(a+\nu)}{\Gamma(a)}
\frac{\Gamma(b+\nu)}{\Gamma(b)}
~, \quad (\nu \neq \mathbb{Z})~,
\label{eq:stationary_Green}
\end{align}
where
\begin{subequations}
\label{eq:def_abc}
\begin{align}
a &= \frac{c}{2} - \frac{i(1+\Omega)}{4\pi T} (\omega-q) 
= \frac{\Deltam}{2} - i\frac{\omega-q}{4\pi \TL}~, \\
b &= \frac{c}{2} - \frac{i(1-\Omega)}{4\pi T} (\omega+q) 
= \frac{\Deltam}{2} - i\frac{\omega+q}{4\pi \TR}~, \\
c &=1-\nu~.
%\label{eq:}
%
\end{align}
\end{subequations}

%%------------------
\subsection{Poles}%\label{sec:}
%%------------------
%\paragraph{Poles:} 
Because $\nu \neq \mathbb{Z}$, the first Gamma functions in the denominator and in the numerator do not diverge. There are 2 types of poles%
\footnote{Gamma functions from the left-movers are called left Gamma functions and poles from the left Gamma functions are called left poles.}:
\begin{subequations}
%\label{eq:}
\begin{align}
\text{left poles: } & a+ \nu = \frac{\Deltap}{2} - i\frac{\omega-q}{4\pi \TL} =-\nlp~, \\ %\quad (\nlp=0,1,\cdots)~,
\text{right poles: } & b+\nu = \frac{\Deltap}{2} - i\frac{\omega+q}{4\pi \TR} = -\nrp~, %\quad (\nrp=0,1,\cdots)~.
%\label{eq:}
%
\end{align}
\end{subequations}
($\nlp,\nrp=0,1,\cdots$)
or
\begin{subequations}
%\label{eq:}
\begin{align}
\omega_L = +q - i(2\pi \TL) (\Deltap + 2\nlp)~,\\
\omega_R = -q - i(2\pi \TR) (\Deltap + 2\nrp)~.
%\label{eq:}
%
\end{align}
\end{subequations}
For fixed real $q$, the poles are evenly-spaced and lie parallel to the negative imaginary axis in the complex $\omega$-plane. As one decreases $\TL$, left poles get closer to one another. In the extreme limit $\TL\to0$, the left poles accumulate to $\omega=q$.

%%------------------
\subsection{Zeros}%\label{sec:}
%%------------------
%\paragraph{Zeros:} 
Similarly, there are 2 types of zeros:
\begin{subequations}
%\label{eq:}
\begin{align}
\text{left zeros: } & a = \frac{\Deltam}{2} - i\frac{\omega-q}{4\pi \TL} = -\nlz~, %\quad (\nlz=0,1,\cdots)~,
\\
\text{right zeros: } & b = \frac{\Deltam}{2} - i\frac{\omega+q}{4\pi \TR} = -\nrz~, %\quad (\nrz=0,1,\cdots)~.
%\label{eq:}
%
\end{align}
\end{subequations}
($\nlz,\nrz=0,1,\cdots$) or
\begin{subequations}
%\label{eq:}
\begin{align}
\omega_L = +q - i(2\pi \TL) (\Deltam + 2\nlz)~,\\
\omega_R = -q - i(2\pi \TR) (\Deltam + 2\nrz)~.
%\label{eq:}
%
\end{align}
\end{subequations}
Again there are an infinite number of zeros.

%%------------------
\subsection{Pole-skip}\label{sec:pole_skip}
%%------------------
%\paragraph{Pole-skip:}
$\Gamma(a)$ and $\Gamma(a+\nu)$ do not diverge simultaneously. Similarly, $\Gamma(b)$ and $\Gamma(b+\nu)$ do not diverge simultaneously. Thus, the remaining possibilities are
\begin{subequations}
\label{eq:pole_skip_stationary}
\begin{align}
& \text{left poles \& right zeros: } 
\nonumber \\
&i\omega = 2\pi \TR\left( \frac{\Deltam}{2}+\nrz \right) + 2\pi \TL\left( \frac{\Deltap}{2}+\nlp \right)~,\\
& iq = 2\pi \TR\left( \frac{\Deltam}{2}+\nrz \right) - 2\pi \TL\left( \frac{\Deltap}{2}+\nlp \right)~,
\\
&\text{right poles \& left zeros: } 
\nonumber \\
&i\omega = 2\pi \TR\left( \frac{\Deltap}{2}+\nrp \right) + 2\pi \TL\left( \frac{\Deltam}{2}+\nlz \right)~,\\
& iq = 2\pi \TR\left( \frac{\Deltap}{2}+\nrp \right) - 2\pi \TL\left( \frac{\Deltam}{2}+\nlz \right)~.
%\label{eq:}
%
\end{align}
\end{subequations}
In the static background, pole-skipping points $\omega$ have a simple form and are given by Matsubara frequencies, but it is no longer the case in the stationary case.
The Green's function consists of two ratios of Gamma functions; one comes from the left-movers and the other comes from right-movers. The pole-skipping occurs when left poles (zeros) coincide with right zeros (poles). Thus, unlike the static case, the pole-skipping points in general depend both on left-temperature $\TL$ and right-temperature $\TR$. Also, note that pole-skipping points depend on the conformal dimensions $\Deltapm$ as well.

When $\TL=\TR$, the $\Deltapm$-dependences disappear from $\omega$, and one gets static BTZ results. The former pole-skipping points become
\begin{subequations}
%\label{eq:}
\begin{align}
i\nw &= 1+\nlp+\nrz~, \\%\quad 
i\nq &= -(\nu+\nlp-\nrz)~, 
%\label{eq:}
%
\end{align}
while the latter become
\begin{align}
i\nw &= 1+\nrp+\nlz~, \\%\quad 
i\nq &= \nu+\nrp-\nlz~.
%\label{eq:}
%
\end{align}
\end{subequations}

The above result uses the BTZ coordinates. In the comoving coordinates \eqref{eq:comoving}, $\omega' =\omega -\Omega q~,  q'=q$, and pole-skipping points \eqref{eq:pole_skip_stationary} are given by Matsubara frequencies like the static case:
\begin{subequations}
%\label{eq:}
\begin{align}
i\nw &= 1+\nlp+\nrz~, \\
i\nw &= 1+\nrp+\nlz~.
%\label{eq:}
%
\end{align}
\end{subequations}
The location of $(iq)$ is the same as the BTZ coordinate results \eqref{eq:pole_skip_stationary}. 

In the comoving coordinates, pole-skipping points take the universal form which is intriguing. However, one needs to be careful. 
%First of all, as is well-known, in the bulk $(2+1)$-dimensions, one can obtain a rotating black hole from a static black hole by a coordinate transformation. They are the same locally. What distinguishes them is the presence of the global boundary condition $x\approx x+2\pi$. So, the metric in the BTZ coordinates with $x\approx x+2\pi$ is physically inequivalent from the one in the comoving coordinates with $x'\approx x'+2\pi$. In this sense, the BTZ coordinates are preferred. In the pole-skipping analysis, one does not really impose the boundary condition and regards $x$ as a noncompact coordinate, but we should keep in mind this special character of the BTZ coordinates. 
From the boundary point of view, the BTZ coordinates correspond to the nonrotating frame whereas the comoving coordinates correspond to the rotating frame. So, it is natural to prefer the BTZ coordinates.

%\paragraph{Slope dependence:}

%%------------------
\subsection{Zero-temperature limit}%\label{sec:}
%%------------------
%\paragraph{Zero-temperature limit:} 
In the extreme limit $\ri\to\ro$ or $\TL\to0$ limit, the arguments of left Gamma functions become all large, and \eq{stationary_Green} reduces to 
\begin{subequations}
%\label{eq:}
\begin{align}
\GR \to& -2\nu (2\pi \TR)^\nu 
%\nonumber  \\
%&\times
\frac{ \Gamma(-\nu) }{ \Gamma(\nu) }
\frac{\Gamma(b+\nu)}{\Gamma(b)} 
\left(\frac{\omega-q}{2i}\right)^\nu~,
%\GR \to& -2\nu e^{-i\pi\nu/2} (2\pi \TR)^\nu 
%\nonumber  \\
%&\times
%\frac{ \Gamma(-\nu) }{ \Gamma(\nu) }
%\frac{\Gamma(b+\nu)}{\Gamma(b)} 
%\left(\frac{\omega-q}{2}\right)^\nu~,
%~, \quad (\nu \neq \mathbb{Z})~,
 \\
b =& \frac{\Deltam}{2} - i\frac{\omega+q}{4\pi \TR}~.
%\label{eq:}
%
\end{align}
\end{subequations}
where we used the asymptotic formula of the Gamma function:
\begin{align}
%
% v1.1
\Gamma(z) \to \sqrt{2\pi} e^{-z}z^{z-1/2}~, \quad (z\to\infty)
\label{eq:gamma_asymptotic}
\end{align}
for $ |\arg{z}| < \pi $. % nonnegative real number $z$. 
One can obtain the result directly by solving the field equation in the extreme background (\appen{extreme_Green}). The left contribution has no poles nor zeros but has the branch point $\omega=q$. 

The Green's function has poles and zeros, but they never coincide. Thus, there is no pole-skipping in the extreme limit. The $T\neq0$ Green's function consists of two ratios of Gamma functions, and each ratio gives an infinite number of poles and zeros. 
%The $T\neq0$ Green's function consists of consists of two sets of Gamma functions, left Gamma functions and right Gamma functions. Each set has Gamma functions both in the numerator and in the denominator and they give an infinite number of poles and zeros. 
The pole-skipping occurs when left poles (zeros) coincide with right zeros (poles). But in the extreme limit, the left Gamma function ratio partly cancels each other and leaves only the power-law behavior in $(\omega-q)$. 

Note that the pole-skipping analysis and the extreme limit do not commute. Namely,
\begin{itemize}
\item 
One should \textit{not} take the extreme limit of nonextreme results \eqref{eq:pole_skip_stationary}. This gives nontrivial results, but they are not correct.
\item
The correct procedure is first to take the extreme limit of the Green's function. Then, carry out the pole-skipping analysis.
\end{itemize}

Further taking the $\TR\to0$ limit,
\begin{align}
\GR \to -2\nu 
\frac{ \Gamma(-\nu) }{ \Gamma(\nu) }
\left( \frac{k^2}{4} \right)^{\nu}~,
%\GR \to -2\nu e^{-i\pi\nu} 
%\frac{ \Gamma(-\nu) }{ \Gamma(\nu) }
%\left( \frac{-k^2}{4} \right)^{\nu}~,
%~, \quad (\nu \neq \mathbb{Z})~,
%\label{eq:}
%
\end{align}
where $k^2:=-\omega^2+q^2$. The result agrees with the pure AdS$_3$ result (Appendix~\ref{sec:AdS_Green} and \ref{sec:complication})%
\footnote{There is a slight complication in the form of the AdS$_3$ Green's function, but it is not relevant to the pole-skipping. See \appen{complication}.}. The Green's function has no pole nor zero but has the branch points at $\omega=\pm q$. 
Because $\GR(k) \propto k^{2\nu}$, the Fourier transformation
\begin{align}
\GR(x)=\int \frac{d^2k}{(2\pi)^2}\, e^{ik\cdot x} \GR(k)
%\label{eq:}
%
\end{align}
implies 
\begin{align}
\GR(x) \propto (-t^2+x^2)^{-\Deltap}~,
%\label{eq:}
%
\end{align}
which is the expected behavior for the Green's function of $\calO$ with the conformal dimension $\Deltap$.

%%%%%%%%%
\section{Integer $\nu$ and extreme pole-skipping}\label{sec:integer_pole_skip}
%%%%%%%%%

So far we assume noninteger $\nu$, and there is no pole-skipping in the extreme limit. The analysis of integer $\nu$ case is more involved, but it is worthwhile to investigate because the pole-skipping occurs even in the extreme limit. 

The conformal dimension of a scalar operator is positive in a unitary CFT$_2$, so we consider the standard quantization. 
%We assume that $\nu$ is a nonnegative integer, which corresponds to the standard quantization \cite{Klebanov:1999tb}. 
The Green's function part of which is relevant to the pole-skipping is given by
\begin{align}
\GR =& (-)^\nu (4\pi^2\TL\TR)^\nu 
%\nonumber \\
%&\times 
\frac{2}{ \Gamma(\nu)^2 }
\frac{\Gamma(a+\nu)}{\Gamma(a)}
\frac{\Gamma(b+\nu)}{\Gamma(b)}
\nonumber \\
&\times 
[\psi(a+\nu)+\psi(b+\nu)]
~, \quad (\nu = \mathbb{Z}^+)~,
\label{eq:nonextreme_green_integer} 
\end{align}
(\appen{integer_Green}) where $\psi(z)$ is a digamma function, and $a,b$ are the same as \eq{def_abc}:
\begin{subequations}
%\label{eq:}
\begin{align}
a &%= \frac{c}{2} - \frac{i(1+\Omega)}{4\pi T} (\omega-q) 
= \frac{\Deltam}{2} - i\frac{\omega-q}{4\pi \TL}~, \\
b &%= \frac{c}{2} - \frac{i(1-\Omega)}{4\pi T} (\omega+q) 
= \frac{\Deltam}{2} - i\frac{\omega+q}{4\pi \TR}~.
%\label{eq:}
%
\end{align}
\end{subequations}

%%------------------
\subsection{Poles}%\label{sec:}
%%------------------
%\paragraph{Poles:}
The arguments of Gamma functions now differ by an integer, so
\begin{align}
\Gamma(a+\nu)=\underbrace{(a+\nu-1)\times \cdots \times a}_{\nu \text{ factors}} \times \Gamma(a)~,
%\label{eq:}
%
\end{align}
and the ratios of the Gamma functions give only zeros.
The digamma functions give poles:
\begin{subequations}
%\label{eq:}
\begin{align}
\text{left poles: } & a+\nu = -\nlp~, %\quad (\nlp=0,1,\cdots)~,
\\
\text{right poles: } & b+ \nu =-\nrp~, %\quad (\nrp=0,1,\cdots)~.
%\label{eq:}
%
\end{align}
\end{subequations}
($\nlp,\nrp=0,1,\cdots$).

%%------------------
\subsection{Zeros}%\label{sec:}
%%------------------
%\paragraph{Zeros:}
From the Gamma functions, 
\begin{subequations}
%\label{eq:}
\begin{align}
\text{left zeros: } & a =-\mlz~, % \quad (\mlz=0,1,\cdots,\nu-1)~,
\\
\text{right zeros: } & b =-\mrz~, % \quad (\mrz=0,1,\cdots,\nu-1)~.
%\label{eq:}
%
\end{align}
\end{subequations}
($\mlz,\mrz=0,1,\cdots,\nu-1$). Unlike the noninteger case, there are only finite number of zeros from the Gamma functions.

In addition, there are zeros when $\psi(a+\nu)+\psi(b+\nu)=0$, but the Gamma functions give only zeros, so there is no pole-skip from these zeros.
%the digamma functions have zeros $\psi(x_n)=0$. 
%No simple formula is known about $x_n$. The first few $x_n$ are given by
%\begin{subequations}
%\label{eq:digamma_zeros}
%\begin{align}
%
%x_0 &\approx 1.46~, 
%x_1 \approx -0.50~, 
%x_2 \approx -1.57~, 
%x_3 \approx -2.61~, 
%\cdots \\
%x_n &= -n + \frac{1}{\pi}\arctan\left(\frac{\pi}{\ln n}\right) + \cdots~.
%x_n &= -n + (\ln n)^{-1} + O[(\ln n)^{-2}]~.
%\label{eq:}
%
%\end{align}
%\end{subequations}

%%------------------
\subsection{Pole-skip}%\label{sec:}
%%------------------
%\paragraph{Pole-skip:}
The pole-skipping points are
\begin{subequations}
\label{eq:pole_skip_stationary_integer}
\begin{align}
&\text{left poles \& right zeros: } 
\nonumber \\
&i\omega = 2\pi \TR\left( \frac{\Deltam}{2}+\mrz \right) + 2\pi \TL\left( \frac{\Deltap}{2}+\nlp \right)~,\\
&iq = 2\pi \TR\left( \frac{\Deltam}{2}+\mrz \right) - 2\pi \TL\left( \frac{\Deltap}{2}+\nlp \right)~,
\\
&\text{right poles \& left zeros: } 
\nonumber \\
&i\omega = 2\pi \TR\left( \frac{\Deltap}{2}+\nrp \right) + 2\pi \TL\left( \frac{\Deltam}{2}+\mlz \right)~,\\
&iq = 2\pi \TR\left( \frac{\Deltap}{2}+\nrp \right) - 2\pi \TL\left( \frac{\Deltam}{2}+\mlz \right)~,
%& (\mlz, \mrz=0,1,\cdots,\nu-1)~.
%\label{eq:}
%
\end{align}
\end{subequations}
%$(\mlz, \mrz=0,1,\cdots,\nu-1)$. 
When $\TL=\TR$, the former gives
\begin{subequations}
%\label{eq:}
\begin{align}
%
%\text{left poles \& right zeros: } 
& i\nw= 1+\nlp+\mrz =: 1+n~, \\
& i\nq= -(\nu+\nlp-\mrz) = -(\nu+ n-2\mrz)~, 
% (\mlz, \mrz=0,1,\cdots,\min(n,\nu-1))~.
\nonumber
%\label{eq:}
%
\end{align}
$[\mlz, \mrz=0,1,\cdots,\min(n,\nu-1)]$ while the latter gives
\begin{align}
%
%\text{right poles \& left zeros: } 
& i\nw= 1+\nrp+\mlz := 1+n~,  \\
& i\nq= \nu+\nrp-\mlz = \nu+n-2\mlz~.
%\label{eq:}
%
\end{align}
\end{subequations}

The pole-skipping formula \eqref{eq:pole_skip_stationary_integer} itself is the same as the noninteger case \eqref{eq:pole_skip_stationary}. The difference is that $(\nlz,\nrz)$ in the noninteger case have no upper bound whereas $(\mlz,\mrz)$ have upper bounds. Consequently, when $n>\nu-1$, some pole-skipping points in noninteger case are ``absent" in the integer case. These points are called ``anomalous points" \cite{Blake:2019otz}.
As an example, consider the static case. When $\Deltap=2~(\nu=1)$, for the second set of pole-skipping, 
\begin{subequations}
%\label{eq:}
\begin{align}
i\nw=1~, \quad 
& i\nq = \nu= 1~, \\
i\nw=2~, \quad 
& i\nq = \nu+1 =2~, \\
\cdots & \nonumber
%\label{eq:}
%
\end{align}
\end{subequations}
At $i\nw_n=n$, there are typically $n$ pole-skipping points for each set, but $(i\nw,i\nq)=(2,0)$ is an anomalous point and is not included%
\footnote{Whether one should include an anomalous point as a pole-skipping point or not depends on how one approaches the pole-skipping point \cite{Natsuume:2019vcv}. But for simplicity, we consider the pole-skipping in a narrow sense and consider the pole-skipping form such as \eq{pole_skip_example}. Namely, we consider the slope dependence of the form $\delta q/\delta\omega$.}.
%For example, when $\Deltap=3~(\nu=2)$, for the second set of pole-skipping, 
%\begin{subequations}
%\label{eq:}
%\begin{align}
%
%i\nw=1~, \quad 
%& i\nq = 2 \\
%i\nw=2~, \quad 
%& i\nq = 3~, 1~,\\
%i\nw=3~, \quad 
%& i\nq = 4~, 2~,\\
%\cdots & \nonumber
%\label{eq:}
%
%\end{align}
%\end{subequations}
%Note that $(i\nw,i\nq)=(3,0)$ is an anomalous point and is not included.

%%------------------
\subsection{Zero-temperature limit}%\label{sec:}
%%------------------
%\paragraph{Zero-temperature limit:} 
In the extreme limit $\TL\to0$, using \eq{gamma_asymptotic} and using the asymptotic formula of the digamma function 
\begin{align}
\psi(z) \to \ln z~, \quad (z\to\infty)
\label{eq:digamma_asymptotic}
\end{align}
%$\psi(z) \to \ln z~(z\to\infty)$ 
for $ |\arg{z}| < \pi $, one obtains 
\begin{subequations}
%\label{eq:}
%\bwt
\begin{align}
\GR \to& (-)^\nu (2\pi \TR)^\nu 
%\nonumber \\
%& \times 
\frac{2}{ \Gamma(\nu)^2 }
\frac{\Gamma(b+\nu)}{\Gamma(b)} 
\left(\frac{\omega-q}{2i}\right)^\nu
 \nonumber \\
& \times 
\left[\ln\frac{\omega-q}{i} + \psi(b+\nu) \right]~,
%\GR \to& (-)^\nu e^{-i\pi\nu/2} (2\pi \TR)^\nu 
%\nonumber \\
%& \times 
%\frac{2}{ \Gamma(\nu)^2 }
%\frac{\Gamma(b+\nu)}{\Gamma(b)} 
%\left(\frac{\omega-q}{2}\right)^\nu
% \nonumber \\
%& \times 
%[\ln(\omega-q)+\psi(b+\nu)]~,
%~, \quad (\nu = \mathbb{Z}^+)~,
\label{eq:extreme_green_integer} \\
b =& \frac{\Deltam}{2} - i\frac{\omega+q}{4\pi \TR}~.
%\label{eq:}
%
\end{align}
%\ewt
\end{subequations}
One can obtain the result directly by solving the field equation in the extreme background (\appen{integer_Green}).
The branch point at $\omega=q$ is replaced by a zero of degree $\nu$. 

The Green's function has poles at
\begin{subequations}
%\label{eq:}
\begin{align}
\text{left (log): } & \omega=q~,  \\
%&& (\text{log}) \\
\text{right (simple pole): } & b+\nu=-\nrp~,  
%&&(\text{simple pole})
%\label{eq:}
%
\end{align}
\end{subequations}
$(\nrp=0,1,\cdots)$ and has zeros at
\begin{subequations}
%\label{eq:}
\begin{align}
\text{left (zero of order $\nu$): } & \omega=q~, \\
\text{right (simple zero): } & b=-\mrz~, 
%\label{eq:}
%
\end{align}
\end{subequations}
$(\mrz=0,1,\cdots,\nu-1)$. In addition, there are zeros when $[\cdots]$ in \eq{extreme_green_integer} vanishes. 

For a generic $\nu$, the pole-skipping does not occur because simple poles and simple zeros remain only in the right-moving sector. However, there is an interesting exception. When $\nu=1$ or $\Deltap=2$, the left zero becomes a simple zero, and it can coincide with the right pole, so the pole-skipping occurs:
\begin{align}
\text{right pole \& left zero: } & 
i\omega = iq= 2\pi \TR (1+\nrp)~.
%\label{eq:}
%
\end{align}
The pole-skipping points are given by right Matsubara frequencies%
\footnote{In order to derive the Green's function \eqref{eq:extreme_green_integer}, we use the formula \eqref{eq:Whittaker_integer}. The formula is actually not valid right at the pole-skipping points, but this is not really a problem because our real interest is the Green's function such as \eq{slope_nu1} which is slightly away from the pole-skipping points. }. 

In this case, the Green's function is given by
\begin{subequations}
%\label{eq:}
\begin{align}
\GR & =
\frac{\omega^2-q^2}{2}
\left[ \ln\frac{\omega-q}{i} + \psi(b+1) \right]
~, \quad (\nu =1)~,
\\
b &= - i\frac{\omega+q}{4\pi \TR}~.
%\label{eq:}
%
\end{align}
\end{subequations}
Near the pole-skipping point, $\psi(b+1)=\psi(-\nrp+\delta b) \sim -1/\delta b$, so the Green's function indeed has the slope-dependence:
\begin{align}
\GR & \sim -(2\pi \TR)^2
2(\nrp+1)
\frac{\delta\omega-\delta q}{\delta\omega+\delta q}~.
\label{eq:slope_nu1}
\end{align}

Going back to \eq{extreme_green_integer}, further taking the $\TR\to0$ limit, 
\begin{align}
\GR \to 
\frac{2}{ \Gamma(\nu)^2 }
\left( \frac{-k^2}{4} \right)^{\nu}
\ln k^2~.
%\ln(-k^2)~.
%~, \quad (\nu = \mathbb{Z}^+)~.
%\label{eq:}
%
\end{align}
The result agrees with the pure AdS$_3$ result (\appen{integer_Green}). 

%%------------------
\subsection{BF bound case}%\label{sec:}
%%------------------
%\paragraph{BF bound:} 
For completeness, let us discuss the $\nu=0$ case where the Breitenlohner-Freedman (BF) bound \cite{Breitenlohner:1982bm} is saturated. In this case, the pole-skipping in the conventional sense does not occur even at finite temperatures. 
%This is the exceptional case in the sense that the pole-skipping never occurs even at finite temperatures. 
The Green's function is given by
\begin{align}
\GR = \frac{1}{2}\psi(a) +\frac{1}{2}\psi(b)
~, \quad (\nu =0)~.
\label{eq:nonextreme_green_BF}
%\label{eq:}
%
\end{align}
Taking the extreme limit $\TL\to0$ gives 
%\begin{align}
%
%G &=
%\left\{ \begin{array}{ll}
%-\ln(\omega-q) -\psi(b)~,  & (\TL,\to0) \\
%-\ln(-k^2)~, & (\TL,\TR\to0)
%\end{array} 
%\right.
%\label{eq:}
%
%\end{align}
\begin{align}
\GR &\to
\frac{1}{2}\ln\frac{\omega-q}{i} + \frac{1}{2}\psi(b)~.
%\label{eq:}
%
\end{align}
Further taking the $\TR\to0$ limit gives
\begin{align}
\GR &\to\frac{1}{2}\ln k^2~.
%\label{eq:}
%
\end{align}
The result agrees with the pure AdS$_3$ result (\appen{integer_Green}). 

Unlike the other cases, the $\nu=0$ Green's function \eqref{eq:nonextreme_green_BF} has no pole-skipping. It has poles at
\begin{subequations}
%\label{eq:}
\begin{align}
\text{left poles: } & a = -\nlp~, %\quad (\nlp=0,1,\cdots)~,
\\
\text{right poles: } & b =-\nrp~, %\quad (\nrp=0,1,\cdots)~,
%\label{eq:}
%
\end{align}
\end{subequations}
$(\nlp,\nrp=0,1,\cdots)$. In addition, there are zeros when $\psi(a)+\psi(b)=0$.

%%%%%%%%%
\section{Power-series expansion method}\label{sec:pole-skipping}
%%%%%%%%%

For the BTZ black hole, the exact Green's function is known so that one can study the pole-skipping directly, but it is instructive to look at power-series expansion method \cite{Grozdanov:2019uhi,Blake:2019otz,Natsuume:2019xcy}. We use the incoming Eddington-Finkelstein (EF) coordinates \eqref{eq:tortoise}. 
%In this section, we briefly review the power-series expansion method of the pole-skipping . 

%%------------------
\subsection{Static case}%\label{sec:}
%%------------------

First, we briefly review the method in static nonextreme backgrounds. 
For simplicity, we set the horizon radius $\ro=1$ in this subsection. 
Typically, the perturbation equation takes the form 
\begin{subequations}
\label{eq:eom}
\begin{align}
0 = \phi''+P(r)\phi'+Q(r)\phi~.
%\label{eq:eom}
%
\end{align}
At nonzero temperature, the horizon $r=1$ is a regular singularity, and $P$ and $Q$ are expanded as
\begin{align}
P &= \frac{P_{-1}}{r-1} + P_0 + P_1(r-1) + \cdots~, \\
Q &= \frac{Q_{-1}}{r-1} + Q_0 + Q_1(r-1) + \cdots~,
%\label{eq:}
%
\end{align}
\end{subequations}
in the EF coordinates. 

%One typically has $P_{-1}=1-i\nw$ and $Q_{-1}=Q_{-1}(\nw,\nq^2)$, where $\nw:=\omega/(2\pi T)$ and $\nq:=q/(2\pi T)$. The field equation takes this form, \eg,  for
%\begin{itemize}
%\item the bulk scalar field,
%\item the bulk Maxwell field (scalar mode and vector mode),
%\item the gravitational perturbations (tensor mode and shear mode).
%\end{itemize}
%We mainly focus on these perturbations, where field equations take the form \eqref{eq:eom}.

The solution can be written as a power series:
\begin{align}
\phi(r) = \sum_{n=0}\, \phi_n\, (r-1)^{n+\lambda}~.
%\label{eq:}
%
\end{align}
Substituting this into the field equation, one obtains the indicial equation at the lowest order: 
\begin{align}
\lambda_1=0~, \quad \lambda_2 =1-P_{-1}~.%= i\nw~.
%\label{eq:}
%
\end{align}
The $\lambda_1$ ($\lambda_2$)-mode represents the incoming (outgoing) mode, and we choose the incoming mode $\lambda_1=0$.
% hereafter. In the incoming EF coordinates, the incoming wave is a Taylor series. The $\lambda_2$-mode is not a Taylor series for a generic $\nw$. 

%The situation changes when $i\nw$ is a nonnegative integer. Then, the $\lambda_2$-mode is also a Taylor series naively. But $\lambda_1$ and $\lambda_2$ differ by an integer. In such a case, the smaller root fails to produce the independent solution since the recursion relation breaks down at some $\phi_n$. Instead, the second solution would contain a $\ln(r-1)$ term and is not regular at $r=1$. 

%However, this log term disappears for special values of $\nq$. Therefore, one has two regular solutions at $i\nw_n=n$ with appropriate $\nq_n$. Such a point is called a ``special point" or a ``pole-skipping point."

The coefficient $\phi_n$ is obtained by a recursion relation. 
At the next order, 
\begin{align}
0 &= Q_{-1} \phi_0 + P_{-1}\phi_1~.
%\label{eq:}
%
\end{align}
Normally, this equation determines $\phi_1$ from $\phi_0$ and gives a unique regular solution. However, when $P_{-1}=Q_{-1}=0$, both $\phi_0$ and $\phi_1$ are free parameters. So, two regular independent solutions are possible. Typically, $P_{-1}=1-i\nw$ for static nonextreme black holes and $Q_{-1}=Q_{-1}(\nw,\nq^2)$. So, this happens at $i\nw=1$ with an appropriate value of $\nq^2$. 
This is the first pole-skipping point.
%Such a point is called a ``special point" or a ``pole-skipping point."

%The former condition gives $M_{12}=P_{-1}=1-i\nw=0$. The latter condition is $M_{11}=Q_{-1}=0$. Since $Q_{-1}$ contains $\nq^2$, there are 2 solutions of $\nq$ and 2 special points. 

Because $P_{-1}=Q_{-1}=0$ at the first pole-skipping point, The horizon $r=1$ changes from a regular singularity to a regular point there. Ref.~\cite{Natsuume:2019xcy} uses this criterion to find $(\nw_1,\nq_1)$. %Also, $\lambda_2=1$ at $\nw_1$, so the extra regular solution is the ``outgoing" solution we did not select previously.

We consider the first pole-skipping point below, but let us briefly discuss higher pole-skipping points for completeness. When $i\nw=n$, the $\lambda_2$-mode is also a Taylor series naively. But $\lambda_1$ and $\lambda_2$ differ by an integer. In such a case, the smaller root actually fails to produce the independent solution since the recursion relation breaks down at some $\phi_n$. Instead, the second solution would contain a $\ln(r-1)$ term and is not regular at $r=1$. 

However, this log term disappears for special values of $\nq$. Therefore, one has two regular solutions at $i\nw_n=n$ with appropriate $\nq_n$. Incidentally, the first pole-skipping point does not have the problem. The horizon becomes a regular point, so two solutions must be regular. 

In order to obtain $(\nw_n, \nq_n)$ systematically, write the perturbation equation in a matrix form \cite{Blake:2019otz}:
\begin{subequations}
%\label{eq:}
\begin{align}
0 & = M\phi \\
& =\begin{pmatrix} 
    M_{11} & M_{12} & 0 & 0 & \cdots \\
    M_{21} & M_{22} & M_{23} & 0 & \cdots \\
    \cdots & \cdots & \cdots & \cdots & \cdots 
  \end{pmatrix}
  \begin{pmatrix} 
    \phi_0 \\ \phi_1 \\ \cdots
  \end{pmatrix}~.
\label{eq:recursion_matrix}
\end{align}
\end{subequations}
Here,
\begin{align}
M_{ij} =a_{ij} i\nw + b_{ij} \nq^2+ c_{ij}~. 
\label{eq:Mij}
\end{align}
In particular, $M_{n,n+1} = n(n-1+P_{-1})=n(n-i\nw)$. The matrix $\calM^{(n)}$ is obtained by keeping the first $n$ rows and $n$ columns of $M$. The pole-skipping points at $i\nw_n=n$ are obtained from
\begin{align}
\det \calM^{(n)}(\nw_n,\nq_n) = 0~.
\label{eq:condition}
\end{align}
This is a degree-$(2n)$ polynomial in $\nq$ since $M_{ij}$ contains $\nq^2$. Thus, typically there are $(2n)$-solutions of $\nq$ and $(2n)$ pole-skipping points at $i\nw=n$.
%For example, consider the first row:
%\begin{align}
%
%0 = M_{11} \phi_0 + M_{12}\phi_1~.
%\label{eq:}
%
%\end{align}
%Normally, this equation determines $\phi_1$ from $\phi_0$. However, when $M_{12}=M_{11}=0$, both $\phi_0$ and $\phi_1$ are free parameters. The former condition gives $M_{12}=P_{-1}=1-i\nw=0$. The latter condition is $M_{11}=Q_{-1}=0$. Since $Q_{-1}$ contains $\nq^2$, there are 2 solutions of $\nq$ and 2 special points. 

%The horizon $r=1$ is a regular singularity, but the horizon becomes a regular point at $(\nw_1,\nq_1)$ because $P_{-1}=Q_{-1}=0$. Ref.~\cite{Natsuume:2019xcy} uses this criterion to find $(\nw_1,\nq_1)$. Also, $\lambda_2=1$ at $\nw_1$, so the extra regular solution is the ``outgoing" solution we did not select previously.

%Similarly, when $M_{23}=\det \calM^{(2)}=0$, $\phi_0$ and $\phi_2$ become free parameters. The former condition gives $M_{23}=2(2-i\nw)=0$. The latter condition is a degree-4 polynomial in $\nq$ since $M_{ij}$ contains $\nq^2$. Thus, there are 4 solutions of $\nq$ and 4 special points. One gets 
%\begin{align}
%
%\det \calM^{(2)} &= Q_{-1}(Q_{-1}+P_0)-P_{-1}Q_0~.
%\label{eq:}
%
%\end{align}

%%------------------
\subsection{Rotating BTZ}%\label{sec:}
%%------------------

Now, consider the rotating BTZ background. In the static case, one typically has $P_{-1}=1-i\nw$, so the first pole-skipping point is given by $i\nw_1=1$ and hence the universality for $i\nw_1$. However, for the rotating BTZ case, \textit{this universality does not hold}. Using the tortoise coordinate \eqref{eq:tortoise}, one obtains
\begin{align}
P_{-1} = 1 - i\nw + i\Omega \nq~.
%\label{eq:}
%
\end{align}
%There are two remarks. 
First, $P_{-1}\neq1-i\nw$ and $P_{-1}$ is $q$-dependent. This explains whey $\nw_1$ does not take the simple form in the rotating background. This is related to the form of the incoming and the outgoing modes in the rotating background (\appen{stationary_Green}). 

Second, in the extreme limit, $P_{-1}$ actually diverges, so the limit is subtle. For example, taking into account $Q_{-1}=0$, one obtains
\begin{subequations}
%\label{eq:}
\begin{align}
i\omega_1 &= \frac{1}{2}(\ro+\ri)(1\mp\nu) + \frac{1}{2}(\ro-\ri)(1\pm\nu) 
\nonumber \\
&= \ro\mp\ri\nu~,\\
iq_1 &=  \frac{1}{2}(\ro+\ri)(1\mp\nu) - \frac{1}{2}(\ro-\ri)(1\pm\nu) 
\nonumber \\
&= \ri\mp\ro\nu~.
%\label{eq:}
%
\end{align}
\end{subequations}
This agrees with the exact Green's function result \eqref{eq:pole_skip_stationary}. If one takes the extreme limit naively, one would get a nontrivial limit, but this should not be trusted. 
In the extreme limit, the horizon becomes degenerate, and the perturbation equation actually has an irregular singularity, so one cannot use Frobenius' method. 

%In this case,
%\begin{subequations}
%\begin{align}
%
%P &= \frac{P_{-2}}{(r-\ro)^2} + \frac{P_{-1}}{r-\ro} + \cdots~, \\
%Q &= \frac{Q_{-2}}{(r-\ro)^2} + \frac{Q_{-1}}{r-\ro} + \cdots~.
%\label{eq:}
%
%\end{align}
%\end{subequations}
Recall the criterion that the horizon becomes a regular point at the first pole-skipping point. For nonextreme black holes,  one has 2 singular terms $P_{-1}=P_{-1}(\nw,\nq)$ and $Q_{-1}=Q_{-1}(\nw,\nq)$. By choosing $\nw$ and $\nq$ appropriately, $P_{-1}$ and $Q_{-1}$ can vanish in general so that the regular singularity becomes a regular point. This fails for extreme black holes, where one generally has more singular terms ($P_{-2}, P_{-1}, Q_{-2}$, and $Q_{-1}$ for the extreme BTZ black hole). In general, one cannot eliminate all terms just by choosing $\nw$ and $\nq$.

However, there is an exception. For the extreme BTZ black hole,
\begin{subequations}
\begin{align}
P &= \frac{ i(q-\omega) }{2(r-\ro)^2} - \frac{ i(q+\omega+4i\ro) }{2\ro(r-\ro)} + \cdots~, \\
Q &= \frac{ -q^2-i\ro(\omega+q)+\ro^2(1-\nu^2) }{ 4\ro^2(r-\ro)^2 } 
\nonumber \\
&+ \frac{ q(q+2i\ro)+\ro^2(1-\nu^2)  }{ 4\ro^3(r-\ro) } + \cdots~.
%\label{eq:}
%
\end{align}
\end{subequations}
By choosing $\omega=q=-i(2\pi\TR) =-2i\ro$ and $\nu^2=1$, the irregular singularity at $r=\ro$ becomes a regular point. So, two regular independent solutions are possible even for the extreme black hole, and \textit{the extreme pole-skipping can occur.} This result agrees with the analysis in previous sections. The power-series expansion method does not work for extreme black holes, but one can find the first pole-skipping point in this way.

%%%%%%%%%
\section{Relation to previous works}\label{sec:previous}
%%%%%%%%%

We studied the scalar Green's function in the rotating BTZ background. Unlike the static case, the pole-skipping points $\omega$ depend both on $\TL$ and $\TR$. 
This has been discussed previously in the context of holographic chaos \cite{Poojary:2018esz,Jahnke:2019gxr,Mezei:2019dfv,Banerjee:2019vff,Liu:2020yaf}. 

In holographic chaos, one is often interested in out-of-time-ordered correlators (OTOC) and computes the Lyapunov exponent $\lambda_L$ and the butterfly velocity $v_B$. According to \cite{Grozdanov:2017ajz}, these quantities can be extracted from the retarded Green's function through the pole-skipping: this was the original motivation of the pole-skipping. In this case, one considers the gravitational sound mode (or the energy-density Green's function.) The scalar field as well as the other bulk fields have pole-skipping points in the lower-half $\omega$-plane. But the gravitational sound mode has a pole-skipping point in the upper-half $\omega$-plane: $\omega_{-1}=+i(2\pi T)$. The quantities $\lambda_L$ and $v_B$ can be extracted by
\begin{align}
\omega_{-1} = i\lambda_L~, \quad q_{-1}=\frac{i\lambda_L}{v_B}~.
%\label{eq:}
%
\end{align}
%v2
Substituting this pole-skipping point into the plane-wave form $e^{-i\omega t+iqx}$, one would get a chaotic behavior.
See, \eg, Refs.~\cite{Gu:2016oyy,Haehl:2018izb,Ramirez:2020qer,Choi:2020tdj} for discussion of the chaotic pole-skipping from field theory point of view. 

%v2
Note that the scalar field as well as the other bulk fields have pole-skipping points in the lower-half $\omega$-plane. So, these pole-skipping points do not seem to indicate chaotic behaviors unlike the gravitational sound mode. But it is still interesting in the following sense. First, the Green's functions are not uniquely determined at pole-skipping points. Second, this is a universal phenomenon which is common to many Green's functions. 

Now, Refs.~\cite{Poojary:2018esz,Jahnke:2019gxr,Mezei:2019dfv,Banerjee:2019vff,Liu:2020yaf} consider the rotating BTZ background and find that $\lambda_L$ depends both on $\TL$ and $\TR$. This implies the corresponding pole-skipping points depend both on $\TL$ and $\TR$. Ref.~\cite{Liu:2020yaf} also points out that $\lambda_L=2\pi T$ in the comoving coordinates. What we find is that the other pole-skipping points (pole-skipping points in the lower-half $\omega$-plane) also depend on both temperatures and depend on $\Deltap$ as well.
%v2
However, we study the scalar Green's function whereas Refs.~\cite{Poojary:2018esz,Jahnke:2019gxr,Mezei:2019dfv,Banerjee:2019vff,Liu:2020yaf} study the energy-density Green's functions. Thus, it is not really possible to compare our results with the ones in Refs.~\cite{Poojary:2018esz,Jahnke:2019gxr,Mezei:2019dfv,Banerjee:2019vff,Liu:2020yaf}.

In the extreme limit, we saw that the pole-skipping does not occur in general, but it does not mean that the pole-skipping never occurs. Indeed, the pole-skipping occurs when $\Deltap=2$. In this case, the pole-skipping points are given by right Matsubara frequencies. We found this phenomenon in the scalar Green's function, but probably this is a generic property of CFT$_2$. 
Various other Green's functions, 
%v2
in particular the energy-density Green's function, 
may also exhibit this phenomenon in principle. In CFT$_2$, the energy-momentum tensor also has $\Deltap=2$, so one probably has an analogous phenomenon there.  
For example, Ref.~\cite{Banerjee:2019vff} studies the OTOC in the extreme BTZ and finds that $\lambda_L$ depends on $\TR$ (in our convention). This is probably an analogous phenomenon in the energy-density Green's function%
%v2
\footnote{The relation between the chaotic behavior in OTOC and the pole-skipping has been explored to some extent for $T\neq0$, but the relation is unclear at zero temperature.
}.

In holographic chaos, OTOC grows exponentially. However, Ref.~\cite{Roberts:2014ifa} points out that OTOC shows a power-law behavior at low temperatures in CFT$_2$. This is called ``slow scrambling," and Ref.~\cite{Craps:2020ahu} studies it from the bulk point of view. It is not known if the pole-skipping has any implications on the slow-scrambling. Our Green's functions show a power-law behavior, \ie,
\begin{align}
\GR(k)\propto k^{2\nu}~, \quad \GR(x)\propto x^{-2\Deltap}~.
%\label{eq:}
%
\end{align}
Of course, this comes from conformal invariance, but the power-law may be related to the power-law in the slow scrambling.
%it is interesting to know if this has anything to do with the slow scrambling.

%%%%%%%%%
\section*{Acknowledgments}%\label{sec:}
%%%%%%%%%

%%%%%%%%%
%\begin{acknowledgments}

%We would like to thank ... for useful discussions. 
This research was supported in part by a Grant-in-Aid for Scientific Research (17K05427) from the Ministry of Education, Culture, Sports, Science and Technology, Japan. 

%\vspace*{0.3cm}
%{\bf Note added}:
%After this paper appeared, Refs.~\cite{Kim:2020url,Sil:2020jhr,Ceplak:2021efc} appeared which also study the pole-skipping. 

%\end{acknowledgments}
%%%%%%%%%

%\newpage
\appendix 

%%%%%%%%%
\section{Preliminaries}%\label{sec:}
%%%%%%%%%

%%------------------
\subsection{Tortoise coordinate}\label{sec:tortoise}
%%------------------

It is worthwhile to recall some basic facts about the tortoise coordinate. Let us consider the following metric:
\begin{subequations}
\label{eq:metric}
\begin{align}
ds^2 &= - Fdt^2+ \frac{dr^2}{F} + r^2 d\vecx^2 \\
&= F(-dt^2+dr_*^2) +\cdots~, \\
%\label{eq:metric} ~.
dr_* &=\frac{dr}{F}~.
\end{align}
\end{subequations}
For simplicity, we set the horizon radius  $\ro=1$. The tortoise coordinate behaves as $r_*\to-\infty$ as $r\to 1$. 

For a nondegenerate horizon, $F\sim (r-1)$, so
\begin{align}
r_* &\sim \int \frac{dr}{r-1} \sim \ln(r-1) \to r-1\sim e^{r_*}~.
%\label{eq:}
%
\end{align}
However, for a degenerate horizon, or for an extreme black hole, $F\sim (r-1)^2$, so
\begin{align}
r_* &\sim \int \frac{dr}{(r-1)^2} \sim \frac{-1}{r-1} \to r-1\sim-\frac{1}{r_*}~.
%\label{eq:}
%
\end{align}

We impose the incoming-wave boundary condition on the horizon. In the tortoise coordinate, the incoming-wave behaves as $e^{-i\omega t}e^{-i\lambda r_*}$. In the BTZ coordinate, it behaves as
\begin{align}
e^{-i\lambda r_*} &\propto 
\left\{ \begin{array}{ll}
(r-1)^{-i\lambda}~, & (\text{nonextreme}) \\
e^{i\lambda/(r-1)}~, & (\text{extreme})
\end{array} 
\right.
\end{align}
In \appen{noninteger_Green}, we use the coordinate $z$ defined by
\begin{subequations}
%\label{eq:}
\begin{align}
z &:= \frac{\ro^2-\ri^2}{r^2-\ri^2} &&\to 1-z\sim r-1~, \quad (\text{nonextreme}) \\
z &:= \frac{\ro^2}{r^2-\ro^2} &&\to z \sim \frac{1}{r-1}~, \quad (\text{extreme})
%\label{eq:}
%
\end{align}
\end{subequations}
Accordingly, 
\begin{align}
e^{-i\lambda r_*} &\propto 
\left\{ \begin{array}{ll}
(1-z)^{-i\lambda}~, & (\text{nonextreme}) \\
e^{i\lambda z}~, & (\text{extreme})
\end{array} 
\right.
%\label{eq:}
%
\end{align}

%%------------------
\subsection{Power-law tail}%\label{sec:}
%%------------------

In the extreme limit, we find the power-law behavior in $k$. This is natural from the CFT point of view. From the bulk point of view, this behavior, the existence of a branch cut in the complex $\omega$-plane, is known as ``power-law tail" \cite{Ching:1995tj}.

The existence of a cut depends on the form of the effective potential in the \textit{tortoise coordinate}. 
Let us consider a massive scalar field. By redefining $\phi$ appropriately $\phi=: G\varphi$, the field equation reduces to the Schr\"{o}dinger form:
\begin{align}
0=\omega^2\varphi + \varphi''-V \varphi~,
%\label{eq:}
%
\end{align}
where $':=\del_*$. For the metric \eqref{eq:metric} in the bulk 4-dimensions, the effective potential $V$ becomes
\begin{align}
V = F \left(m^2+\frac{F'}{r} \right)~.
%\label{eq:}
%
\end{align}
What is important is the overall factor $F$. 

Consider the potential near the horizon. For a nondegenerate horizon,
\begin{align}
V &\propto r-1 \propto e^{r_*}~, \quad (r\to1)~,
%\label{eq:}
%
\end{align}
so the potential decays exponentially. On the other hand, for a degenerate horizon or an extreme black hole, 
\begin{align}
V &\propto (r-1)^2 \propto \frac{1}{r_*^2}~, \quad (r\to1)~,
%\label{eq:}
%
\end{align}
so the potential only decays as a power law. The cut is a general feature if $V$ tends to zero slower than an exponential, so a cut should exist for extreme black holes from the near-horizon behavior. 

The power-law tail is however not restricted to extreme black holes. Rather, it is often discussed for asymptotic flat nonextreme black holes. For nonextreme black holes, $V$ decays exponentially near the horizon, but $V$ can decay as a power-law asymptotically. 
For the Schwarzschild black hole, $F=1-r^{-1}$, so $r\sim r_*$ asymptotically. The potential then behaves as
\begin{align}
V \sim m^2-\frac{m^2}{r} \sim  m^2-\frac{m^2}{r_*}~, \quad (r\to\infty)~.
%\label{eq:}
%
\end{align}
For AdS black holes, the effective potential diverges asymptotically, so a cut does not come from the asymptotic behavior and comes only from the near-horizon behavior.

Finally, the effective potential of a rotating black hole is very similar. If the metric takes the form 
\begin{align}
ds^2 &= - Fdt^2+ \frac{dr^2}{F} + r^2 (dx+N^x dt)^2~,
%\label{eq:metric} 
%
\end{align}
the difference is just $\omega^2 \to (\omega+q N^x)^2$ in the Schr\"{o}dinger problem, where $N^x$ is the shift function.
% A similar discussion should be possible, but it is not extensively studied.

%%%%%%%%%
\section{Scalar Green's functions (noninteger $\nu$)}\label{sec:noninteger_Green}
%%%%%%%%%

We derive the retarded Green's functions for a scalar field in the following backgrounds: %BTZ black holes and in the pure AdS$_3$. 
\begin{itemize}
\item pure AdS$_3$
\item (nonextreme) BTZ black holes
\item (extreme) BTZ black holes
\end{itemize}
The Green's functions for the first two backgrounds were originally derived in Refs.~\cite{Birmingham:2001pj,Son:2002sd} and were reproduced in many literature, but we rederive them for completeness. 
%v2
On the other hand, the Green's function for the extreme background has not been derived as far as we are aware.
We assume $\nu\neq\mathbb{Z}$ in this appendix. The $\nu=\mathbb{Z}$ case is discussed separately in \appen{integer_Green}. 

%%------------------
\subsection{Pure AdS$_3$}\label{sec:AdS_Green}
%%------------------

%The metric is given by
%\begin{subequations}
%\label{eq:}
%\begin{align}
%
%ds^2 &= -r^2 dt^2 + \frac{dr^2}{r^2} +r^2 dx^2~, \\
%&=\frac{1}{u^2}(-dt^2+du^2+dx^2)~,
%\label{eq:}
%
%\end{align}
%\end{subequations}
%where we set the AdS radius $L=1$ and $u:=1/r$. 
%The metric 
%The $T\to0$ limit of the static BTZ black hole 
When $\ro=\ri=0$, the BTZ black hole reduces to the pure AdS$_3$ if $x$ is noncompact or reduces to the ``zero-mass" black hole if $x$ is compact. 
We consider the perturbation of the form $\phi(u)e^{-i\omega t+iqx}$ where $u:=1/r$. We first assume that $\omega>0$ and $k^2=-\omega^2+q^2$ is timelike, $k^2<0$. The scalar field equation $(\nabla^2-m^2)\phi=0$ becomes
\begin{align}
0= \phi''-\frac{1}{u}\phi'+ \left(p^2-\frac{1-\nu^2}{u^2}\right)\phi~, 
%\label{eq:}
%
\end{align}
where $p := \sqrt{-k^2} = \sqrt{\omega^2-q^2}$ and $\nu := \sqrt{1+m^2}$. 
%Asymptotically, the solution behaves as 
%\begin{subequations}
%\label{eq:}
%\begin{align}
%
%\phi &\sim A u^{\Deltam} + B u^{\Deltap}~,
%\quad (u\to0)~,\\
%\Delta_\pm &=1\pm \nu~.
%\label{eq:}
%
%\end{align}
%\end{subequations}
%According to the standard AdS/CFT dictionary, the slow falloff $A$ represents the source of the operator $\calO$, and the fast falloff $B$ represents the response to $\calO$:
%\begin{align}
%
%\bra\calO\ket = 2\nu B~.
%\label{eq:}
%
%\end{align}
%Then, the retarded Green's function $\GR$ is given by
%\begin{align}
%
%\GR = -\frac{\delta\bra\calO\ket}{\delta A} = -(2\nu)\frac{B}{A}~.
%\label{eq:}
%
%\end{align}

Defining $z:=pu$ and setting $\phi=z f$, the field equation reduces to 
\begin{align}
0 = \del_z^2 f+\frac{1}{z}\del_z f + \left( 1-\frac{\nu^2}{z^2} \right) f~,
%\label{eq:}
%
\end{align}
and the solutions are given by Bessel functions. The differential equation has a regular singularity at $z=0$ and an irregular singularity at $z=\infty$. It is convenient to use the Hankel functions $H_{\nu}^{(1,2)}$ % $H_{\nu}^{(1,2)}:=J_{\nu}\pm iY_{\nu}$ 
because they take the plane-wave form:
\begin{align}
H_{\nu}^{(1,2)} (z) \sim \sqrt{\frac{2}{\pi z}} e^{\pm iz \mp \pi i(2\nu+1)/4}~, \quad(z\to\infty)~,
%\label{eq:}
%
\end{align}
% to be confirmed
for $ |\arg{z}| < \pi $. 
The pure AdS$_3$ is not a black hole, but it has the ``Poincare horizon" at $z\to\infty$ and one imposes the incoming-wave boundary condition there; the boundary condition picks up the solution
\begin{align}
f \propto H_{\nu}^{(1)} (p u)~.
%\label{eq:}
%
\end{align}
Then, the solution takes the form of the incoming-wave:
\begin{align}
\phi e^{-i\omega t} = H_{\nu}^{(1)} e^{-i\omega t} \sim e^{-i\omega t+ ipu}~.
%\label{eq:}
%
\end{align}

The Hankel functions have the following expansion:
%In order to extract the asymptotic behavior,
\begin{subequations}
%\label{eq:}
\begin{align}
H_\nu^{(1,2)}
=& \frac{\mp i}{\sin\nu\pi}(J_{-\nu}-e^{\mp i\pi\nu}J_\nu)~, \quad (\nu \neq \mathbb{Z}) 
\label{eq:hankel_noninteger} \\
\sim& \frac{\pm i}{\sin\nu\pi} \biggl[ - \frac{1}{\Gamma(1-\nu)}\left(\frac{z}{2}\right)^{-\nu} + \cdots
\nonumber \\
&+ e^{\mp i\pi\nu} \frac{1}{\Gamma(1+\nu)}\left(\frac{z}{2}\right)^\nu +\cdots \biggr]~.
\label{eq:hankel_asymptotic}
\end{align}
\end{subequations}
The Green's function is then given by
\begin{align}
\GR = -2\nu e^{-i\pi\nu} \frac{ \Gamma(-\nu) }{ \Gamma(\nu) }
\left( \frac{-k^2}{4} \right)^{\nu}~.
%~, \quad (\nu \neq \mathbb{Z})~.
%\label{eq:}
%
\end{align}
When $\nu=\mathbb{Z}$, the definition of the Hankel function \eqref{eq:hankel_noninteger} is ill-defined and should be replaced by \eq{hankel_integer}.

%%------------------
\subsection{A remark}\label{sec:complication}
%%------------------

It is customary to write the AdS$_3$ Green's function using the function $\text{sgn}(\omega)$. See Ref.~\cite{Son:2002sd} for the details. 

%When we obtain the Green's function for AdS$_3$, 
So far we assume that $\omega>0$ and $k^2<0$. In this case, $H_\nu^{(1)}$ is the incoming-wave. But when $\omega<0$ and $k^2<0$, one chooses $H_\nu^{(2)}$ as the incoming-wave. Then, the Green's function is given by
\begin{align}
\GR &= -2\nu e^{-i\pi\nu\,\text{sgn}(\omega)} \frac{ \Gamma(-\nu) }{ \Gamma(\nu) }
\left( \frac{-k^2}{4} \right)^{\nu}~,
~ (k^2<0)~. 
\label{eq:AdS_Green_timelike}
\end{align}
When $k^2>0$, one chooses the regularity condition at the horizon, and the Green's function becomes
\begin{align}
\GR &= -2\nu \frac{ \Gamma(-\nu) }{ \Gamma(\nu) }
\left( \frac{k^2}{4} \right)^{\nu}~,
~ (k^2>0)~.
\label{eq:AdS_Green_spacelike}
\end{align}

We obtain the Green's functions for integer $\nu$ in \appen{integer_Green}, but there is a similar complication.
Also, for the extreme BTZ black hole, the issue of the incoming-wave does not arise, but there is a similar complication. 

The Green's function using the $\text{sgn}$ function is useful (\eg, to extract the real and imaginary parts), but it is most suitable for real $\omega$ and $q$. Also, the function is not really relevant to the pole-skipping. 
%The presence of $\text{sgn}\,\omega$ is important for consistency, but we ignore the factor in the text since it is not relevant to the pole-skipping.
To avoid the complication, we write the AdS$_3$ Green's function as
\begin{align}
\GR &= -2\nu \frac{ \Gamma(-\nu) }{ \Gamma(\nu) }
\left( \frac{k^2}{4} \right)^{\nu}~,
%~ (k^2>0)~.
\label{eq:AdS_Green}
\end{align}
both for $k^2>0$ and for $k^2<0$. Here, 
%Then, \eq{AdS_Green_timelike} is obtained as follows:
\begin{itemize}
\item $k^2$ means %Replace
\begin{subequations}
\label{eq:k^2_rule}
\begin{align}
k^2 \to \frac{\omega-q}{i}\frac{\omega+q}{i},
%\label{eq:}
%
\end{align}
\item 
and we take
\begin{align}
\left| \arg{\frac{\omega\pm q}{i}}\right| <\pi~.
\label{eq:argument}
\end{align}
\end{subequations}
\end{itemize}

% v1.1
For real $\omega$ and $q$, \eq{AdS_Green} reduces to Eqs.~\eqref{eq:AdS_Green_timelike} and \eqref{eq:AdS_Green_spacelike}.
The procedure implies
\begin{align}
\frac{\omega-q}{i} =|\omega-q| e^{-i\pi\,\text{sgn}(\omega-q)/2}~.
%\label{eq:}
%
\end{align}
Note that
\begin{subequations}
%\label{eq:}
\begin{align}
k^2<0: &|\omega|>|q| \to \text{sgn}(\omega-q) = \text{sgn}(\omega)~, \\
k^2>0: &|\omega|<|q| \to \text{sgn}(\omega-q) = -\text{sgn}(q)~.
%\label{eq:}
%
\end{align}
\end{subequations}
%For timelike $k^2$, $|\omega|>|q|$, so $\text{sgn}(\omega-q) = \text{sgn}(\omega)$.
%For spacelike $k^2$, $|\omega|<|q|$, so $\text{sgn}(\omega-q) =-\text{sgn}(q)$.
Then,
\begin{subequations}
%\label{eq:}
\begin{align}
k^2 &\to |\omega^2-q^2| e^{-i\pi\{ \text{sgn}(\omega-q)+\text{sgn}(\omega+q) \}/2} \\
&= \left\{ \begin{array}{ll} 
|\omega^2-q^2| e^{-i\pi\,\text{sgn}(\omega)} & (k^2<0) \\
|\omega^2-q^2| & (k^2>0)
\end{array}
\right.
%\label{eq:}
%
\end{align}
\end{subequations}
Thus, \eq{AdS_Green} reduces to Eqs.~\eqref{eq:AdS_Green_timelike} and \eqref{eq:AdS_Green_spacelike}.

Here, we consider real $(\omega,q)$ only, but we are also interested in complex $(\omega,q)$.
Also, the same convention is used for integer $\nu$ results and for extreme black hole results. 

One can understand the above procedure starting from the finite-temperature Green's function. The above rules come from taking the zero-temperature limit. In particular, \eq{argument} comes from the asymptotic formula of the Gamma function \eqref{eq:gamma_asymptotic}.

At finite temperatures, there are lines of poles parallel to the negative imaginary axis for real $q$. In the zero-temperature limit, the lines of poles are replaced by branch cuts from $\omega=\pm q$. \eq{argument} implies 
% v1.1: corrected
$-\pi/2<\arg{(\omega\pm q)}<3\pi/2$, so one can avoid the branch cuts.

%%------------------
\subsection{Rotating BTZ}\label{sec:stationary_Green}
%%------------------

Ref.~\cite{Son:2002sd} obtains the scalar Green's function for the nonextreme BTZ black hole in an elegant manner, but we derive it using a brute-force method. 

For the nonextreme BTZ black hole, we use the radial coordinate $z$:
\begin{align}
z:= \frac{\ro^2-\ri^2}{r^2-\ri^2}~.
\label{eq:radial_nonextreme}
\end{align}
The asymptotic infinity is located at $z=0$ and the outer horizon is located at $z=1$.
The scalar field equation then becomes
\begin{subequations}
%\label{eq:}
\begin{align}
0 &= \left[ h\del_z(h\del_z) + \frac{ (\omega-\Omega q)^2-(\Omega\omega-q)h }{ (4\pi T)^2 z }
- \frac{m^2h}{4z^2} \right]\phi~, \\
h &=1-z~.
%\label{eq:}
%
\end{align}
\end{subequations}

Asymptotically $z\to0$, the field equation behaves as
\begin{align}
0\sim \del_z^2 \phi -\frac{\nu^2-1}{4z^2} \phi~,
%\label{eq:}
%
\end{align}
so the solution behaves as $\phi \sim z^{\Deltapm}/2$, where $\Deltapm=1\pm\nu$.

Near the horizon $z\to1$, the field equation behaves as 
\begin{subequations}
%\label{eq:}
\begin{align}
0 &\sim h\del_z(h\del_z \phi) +\lambda^2 \phi~, \\
\lambda &:=\frac{\omega-\Omega q}{4\pi T}~,
%\label{eq:}
%
\end{align}
\end{subequations}
so the near-horizon solution is 
\begin{align}
\phi \sim (1-z)^{\pm i\lambda}~.
\label{eq:incoming_stationary}
\end{align}
The incoming-wave is $\phi \sim (1-z)^{-i\lambda}$. In the static background, $\lambda=\omega/(4\pi T)$, which is a familiar result. But in the rotating background, this does not hold in general, and the exponent $\lambda$ is $q$-dependent as well. Also, one can use the comoving coordinates \eqref{eq:comoving} where the incoming-wave takes the same form as the static case.

Taking into account these behaviors, set the ansatz 
\begin{align}
\phi = z^{(1-\nu)/2}(1-z)^{-i\lambda} f(z)~.
\label{eq:ansatz_stationary}
\end{align}
Then, the field equation becomes the hypergeometric differential equation:
\begin{align}
0 = z(1-z)\del_z^2 f + \{ c-(1+a+b)z\} \del_z f - ab f~, 
%\label{eq:}
%
\end{align}
where
\begin{subequations}
%\label{eq:}
\begin{align}
a &= \frac{c}{2} - \frac{i(1+\Omega)}{4\pi T} (\omega-q) 
= \frac{\Deltam}{2} - i\frac{\omega-q}{4\pi \TL}~, \\
b &= \frac{c}{2} - \frac{i(1-\Omega)}{4\pi T} (\omega+q) 
= \frac{\Deltam}{2} - i\frac{\omega+q}{4\pi \TR}~, \\
c &=1-\nu~.
%\label{eq:}
%
\end{align}
\end{subequations}
Note $a+b=c-2i\lambda$. The differential equation has 3 regular singularities at $z=0,1$, and $\infty$. The solutions are given by hypergeometric functions. 

Asymptotically $z\to 0$, the independent solutions are
\begin{align*}
& {}_2F_1(a,b,c;z) \sim 1~, \\ %\quad
& z^{1-c}{}_2F_1(1+a-c,1+b-c,2-c;z) \sim z^{\nu}~.
%\label{eq:}
%
\end{align*}
Combined with the ansatz \eqref{eq:ansatz_stationary}, the former gives the slow falloff (source) and the latter gives the fast falloff (response). The hypergeometric function ${}_2F_1(a,b,c;z)$ is ill-defined when $c$ is a nonpositive integer which corresponds to $\nu=1,2,\cdots$. We assume $\nu\neq\mathbb{Z}$. 

Near the horizon $z\to1$, the independent solutions are
\begin{align*}
& {}_2F_1(a,b,1+a+b-c;1-z)~, \\%\quad
& (1-z)^{c-a-b}{}_2F_1(c-a,c-b,1+c-a-b;1-z)~.
%\label{eq:}
%
\end{align*}
The former gives the incoming-wave whereas the latter gives the outgoing wave. We impose the incoming-wave boundary condition. In order to extract the asymptotic behavior, it is convenient to use the formula:
% corrected
\begin{align}
\lefteqn{ {}_2F_1(\alpha,\beta,\gamma ;z) = 
\frac{ \Gamma(\gamma)\Gamma(\alpha+\beta-\gamma) }{ \Gamma(\alpha)\Gamma(\beta)} 
}
\label{eq:Gauss_transf} \\
&\times (1-z)^{\gamma-\alpha-\beta} {}_2F_1(\gamma-\alpha, \gamma-\beta, 1+\gamma-\alpha-\beta ;1-z)
\nonumber \\
&+\frac{ \Gamma(\gamma)\Gamma(\gamma-\alpha-\beta) }{ \Gamma(\gamma-\alpha)\Gamma(\gamma-\beta)} 
{}_2F_1(\alpha,\beta,1+\alpha+\beta-\gamma ;1-z)~.
\nonumber 
\end{align}
Then, the incoming wave is written as
\begin{align}
&{}_2F_1(a,b,a+b+\nu;1-z) \propto 
{}_2F_1(a,b,1-\nu;z)  
\nonumber \\
&+ \frac{ \Gamma(-\nu) }{ \Gamma(\nu) }\frac{ \Gamma(a+\nu) }{ \Gamma(a) }\frac{ \Gamma(b+\nu) }{ \Gamma(b) } 
\nonumber \\
& \times 
% corrected
z^{\nu} {}_2F_1(a+\nu,b+\nu,\nu+1;z)~.
%&{}_2F_1(a,b,1+a+b-c;1-z) \propto 
%{}_2F_1(a,b,c;z)  \nonumber \\
%&+ \frac{ \Gamma(c-1) }{ \Gamma(1-c) }\frac{ \Gamma(1-c+a) }{ \Gamma(a) }\frac{ \Gamma(1-c+b) }{ \Gamma(b) } (1-z)^{1-c} {}_2F_1(1+a-c,1+b-c,2-c;z)~.
%\label{eq:}
%
\end{align}

When a solution behaves as 
\begin{align}
\phi &\sim A r^{-\Deltam} + B r^{-\Deltap}~,
%\label{eq:}
%
\end{align}
%the expectation value of the operator $\calO$ which is dual to the scalar field is given by
%\begin{align}
%
%\bra\calO\ket = 2\nu B~,
%\label{eq:}
%
%\end{align}
%and 
the retarded Green's function $\GR$ is given by
\begin{align}
\GR = -(2\nu)\frac{B}{A}~.
%\label{eq:}
%
\end{align}
But note that we use the coordinate $z$ which behaves as
\begin{align}
r^2 \sim \frac{\ro^2-\ri^2}{z}~, \quad (r\to\infty)~.
%\label{eq:}
%
\end{align}
Accordingly, the falloffs $A$ and $B$ should be extracted from
\begin{align}
\phi \sim 
A \left( \frac{z}{\ro^2 \!-\! \ri^2} \right)^{\Deltam/2} 
+ B \left( \frac{z}{\ro^2 \!-\! \ri^2} \right)^{\Deltap/2}~,
\quad (z\to0)~.
%\label{eq:}
%
\end{align}
Then, the Green's function is given by
\begin{align}
\GR = -2\nu(4\pi^2\TL\TR)^\nu 
\frac{ \Gamma(-\nu) }{ \Gamma(\nu) }
\frac{\Gamma(a+\nu)}{\Gamma(a)}
\frac{\Gamma(b+\nu)}{\Gamma(b)}~.
% \quad(\nu\neq\mathbb{Z})
\label{eq:nonextreme_green}
\end{align}

%%------------------
\subsection{Alternative method}%\label{sec:}
%%------------------

%v2
In previous subsection, we obtained the Green's function by solving the field equation directly in the rotating background. However, one can obtain the same result by solving the field equation in the static background and by utilizing a coordinate transformation. 

In the bulk $(2+1)$-dimensions, one can obtain a rotating background from a static background by a coordinate transformation. So, consider the Lorentz boost on the boundary coordinates:
\begin{subequations}
\label{eq:boost}
\begin{align}
d\hatt &= \gamma(dt-\Omega\,dx)~, \\
d\hx &= \gamma(dx-\Omega\,dt)~,
%\label{eq:}
%
\end{align}
\end{subequations}
where $\gamma := 1/\sqrt{1-\Omega^2}$. Here, the coordinate system $(t,x)$ represents the rotating metric. In the coordinate system $(\hatt,\hx)$, the metric takes the form of the static metric:
\begin{align}
ds^2= \hr_+^2\frac{-hd\hatt^2+d\hx^2}{z} + \frac{dz^2}{4hz^2}~.
%\label{eq:}
%
\end{align}
Here, the horizon radius $\hr_+$ is given by
\begin{align}
\hr_+=\frac{\ro}{\gamma}~,
%\label{eq:}
%
\end{align}
and the Hawking temperature $2\pi\hT=\hr_+$. In momentum space, 
\begin{subequations}
%\label{eq:}
\begin{align}
\homega &= \gamma(\omega - \Omega\,q)~, \\
\hq &= \gamma(q - \Omega\,\omega)~,
%\label{eq:}
%
\end{align}
or
\begin{align}
\frac{\homega\pm \hq}{2\pi\hT} = \frac{(1\mp\Omega)(\omega \pm q)}{2\pi T}~.
%\label{eq:}
%
\end{align}
\end{subequations}

Thus, one can first solve the field equation in the static background $(\hatt,\hx)$. Then, transform the results into the rotating background using the above formulae.
For example, the incoming wave behaves as 
\begin{align}
(1-z)^{-i\homega/(4\pi\hT)}
%\label{eq:}
%
\end{align}
in the static background.
Under the Lorentz boost, the exponent becomes
\begin{align}
-i\frac{\homega}{4\pi\hT} = -i\frac{\omega-\Omega q}{4\pi T}
%\label{eq:}
%
\end{align}
as expected from \eq{incoming_stationary}.

The Lorentz boost transforms the rotating metric to the static metric, but this does not imply that the rotating metric is equivalent to the static metric. They are the same locally, but the presence of the global boundary condition distinguishes them. The metric with $x\approx x+2\pi$ is physically inequivalent from the one with $\hx\approx \hx+2\pi$. 

%Note that the Lorentz boost does not imply that the rotating metric is equivalent to the static metric because one actually has to impose the periodic boundary condition on $x$ or $\hx$ (\sect{pole_skip}). 
Also, the comoving coordinate \eqref{eq:comoving} is given by the {\it Galilean} boost instead of the Lorentz boost \eqref{eq:boost}. Namely, the effect of the black hole rotation cannot be completely eliminated by the Galilean boost.

%%------------------
\subsection{Extreme BTZ}\label{sec:extreme_Green}
%%------------------

In the extreme limit, one cannot use the radial coordinate \eqref{eq:radial_nonextreme}. Instead, we use the radial coordinate $z$:
\begin{align}
z:= \frac{\ro^2}{r^2-\ro^2}~.
%\label{eq:}
%
\end{align}
The asymptotic infinity is located at $z=0$ and the horizon is located at $z=\infty$.

Further introducing 
\begin{align}
\zeta:= -\frac{i(\omega-q)}{2\pi \TR} z  = -2i\lambda z~,
%\label{eq:}
%
\end{align}
the scalar field equation reduces to the Whittaker differential equation: 
\begin{subequations}
%\label{eq:}
\begin{align}
0 &= \del_\zeta^2 \phi + \left( -\frac{1}{4} + \frac{\kappa}{\zeta} - \frac{(\nu/2)^2 - 1/4}{\zeta^2} \right)\phi~, 
\label{eq:whittaker}\\
\kappa &:= \frac{i(\omega+q)}{4\pi \TR}~.
%\label{eq:}
%
\end{align}
\end{subequations}
The differential equation has a regular singularity at $\zeta=0$ and an irregular singularity at the degenerate horizon $\zeta\to\infty$. So, one cannot use the power-series method in \sect{pole-skipping}.

Asymptotically $\zeta\to0$, the field equation behaves as
\begin{align}
0\sim \del_\zeta^2 \phi -\frac{\nu^2-1}{4\zeta^2}\phi~,
%\label{eq:}
%
\end{align}
so the solution behaves as $\phi \sim \zeta^{\Deltapm}/2$.

Near the horizon $\zeta\to-i\infty$, the field equation behaves as 
\begin{align}
0 &\sim \del_\zeta^2\phi - \frac{1}{4}\phi~,
%\label{eq:}
%
\end{align}
so the near-horizon solution is 
\begin{align}
\phi \sim e^{\pm\zeta/2} \sim e^{\mp i\lambda z}~.
%\label{eq:}
%
\end{align}
The incoming-wave is
\begin{align}
\phi \sim e^{-\zeta/2} \sim e^{i\lambda z}~.
%\label{eq:}
%
\end{align}

The form of the incoming-wave is different from the nonextreme case. This is because the tortoise coordinate $r_*$ takes a different form in the extreme case (\appen{tortoise}). The coordinate $z$ itself is proportional to the tortoise coordinate. Also, note that the exponent $\lambda$ is $q$-dependent just like the nonextreme case. 

The Whittaker functions 
\begin{align}
W_{\kappa,\nu/2}(\zeta), W_{-\kappa,\nu/2}(-\zeta), M_{\kappa,\nu/2}(\zeta), M_{\kappa,-\nu/2}(\zeta)
%\label{eq:}
%
\end{align}
are the solutions of the Whittaker differential equation \eqref{eq:whittaker}. The function $M_{\kappa,\nu/2}$ is ill-defined when $\nu$ is a negative integer. We assume $\nu \neq \mathbb{Z}$. Then, $W_{\kappa,\nu/2}(\zeta)$ and $W_{-\kappa,\nu/2}(-\zeta)$ (or $M_{\kappa,\nu/2}(\zeta)$ and $M_{\kappa,-\nu/2}(\zeta)$) are independent solutions. Near the horizon, it is convenient to choose $W_{\kappa,\nu/2}(\zeta)$ and $W_{-\kappa,\nu/2}(-\zeta)$ because they take the plane-wave form:
\begin{align}
%
% v1.1
W_{\kappa,\nu/2}(\zeta) \sim e^{-\zeta/2} \zeta^\kappa~, \quad (\zeta\to\infty)
%\label{eq:}
%
\end{align}
% to be confirmed
for $ |\arg{\zeta}| < \pi $.
Asymptotically $\zeta\to0$, it is convenient to choose $M_{\kappa,\nu/2}(\zeta)$ and $M_{\kappa,-\nu/2}(\zeta)$ because they admit power-series solutions:
\begin{align}
M_{\kappa,\nu/2}(\zeta) \sim \zeta^{(1+\nu)/2} e^{-\zeta/2}~, \quad (\zeta\to0)~.
%\label{eq:}
%
\end{align}
From the incoming-wave boundary condition, 
\begin{align}
\phi \propto W_{\kappa,\nu/2}(-2i\lambda z) \propto e^{i\lambda z}~.
%\label{eq:}
%
\end{align}
%Then, the solution takes the incoming-wave:
%\begin{align}
%
%\phi e^{-i\omega t} \sim e^{-i\omega t+i\lambda z}~, \quad (z\to\infty)~. 
%\label{eq:}
%
%\end{align}

In order to extract the asymptotic behavior, one can use the relation between Whittaker functions:
\begin{align}
W_{\kappa,\nu/2}(\zeta) &= 
\frac{ \Gamma(-\nu) }{ \Gamma(\half-\half\nu-\kappa) } M_{\kappa,\nu/2}(\zeta) 
\nonumber \\
&+ \frac{ \Gamma(\nu) }{ \Gamma(\half+\half\nu-\kappa) } M_{\kappa,-\nu/2}(\zeta)~.
\label{eq:Whittaker_transf}
\end{align}
$M_{\kappa,-\nu/2} \sim \zeta^{(1-\nu)/2}$ represents the slow falloff, and $M_{\kappa,\nu/2} \sim \zeta^{(1+\nu)/2}$ represents the fast falloff. Taking the ratio of their coefficients, one obtains
\begin{subequations}
%\label{eq:}
\begin{align}
\GR &=-2\nu (2\pi \TR)^\nu 
\frac{ \Gamma(-\nu) }{ \Gamma(\nu) }
\frac{\Gamma(b+\nu)}{\Gamma(b)} 
\left(\frac{\omega-q}{2i}\right)^\nu~,
%\GR &=-2\nu e^{-i\pi\nu/2} (2\pi \TR)^\nu 
%\frac{ \Gamma(-\nu) }{ \Gamma(\nu) }
%\frac{\Gamma(b+\nu)}{\Gamma(b)} 
%\left(\frac{\omega-q}{2}\right)^\nu~,
%~, \quad (\nu \neq \mathbb{Z})~.
\label{eq:extreme_green} \\
b &= \frac{\Deltam}{2} - i\frac{\omega+q}{4\pi \TR}~,
\end{align}
\end{subequations}
where we take into account the fact that the coordinate $z$ behaves as $r^2\sim \ro^2/z~(r\to\infty)$. So, just like the nonextreme case, the falloffs $A$ and $B$ should be extracted from
\begin{align}
\phi \sim A \left(\frac{\sqrt{z}}{\ro}\right)^{\Deltam} + B \left(\frac{\sqrt{z}}{\ro}\right)^{\Deltap}~, 
\quad (z\to0)~.
%\label{eq:}
%
\end{align}
% v1.1
Also, the result should be understood under the convention \eqref{eq:k^2_rule}.

%%%%%%%%%
\section{Scalar Green's functions (integer $\nu$)}\label{sec:integer_Green}
%%%%%%%%%

%%------------------
\subsection{AdS/CFT dictionary for integer $\nu$}%\label{sec:}
%%------------------

For integer $\nu$, the Green's function and its pole-skipping are more involved. So far, we construct solutions using special functions, and they can be expanded as a power series asymptotically:
\begin{align}
\phi \sim A u^{1-\nu}(1+\cdots) + B u^{1+\nu}(1+\cdots)~,
%\quad (u\to0)~,
%\label{eq:}
%
\end{align}
where $u=1/r$. However, when $\nu=\mathbb{Z}$, the exponents differ by an integer. In such a case, one cannot construct independent solutions by the power-series expansion in general, and one expects that log terms appear.
%two solutions are not independent, and one expects that the independent solution includes log terms. 
The dictionary is as follows:
\begin{itemize}
\item When $\nu>0$,
\begin{align}
\phi \sim& A u^{1-\nu}\left\{1+\cdots +O(u^{2\nu}\ln u)+\cdots\right\} 
\nonumber \\
&+ B u^{1+\nu}(1+\cdots)~,
%\quad (u\to0)~.
%\label{eq:}
%
\end{align}
The Green's function is given by
\begin{align}
\GR = -2\nu \frac{B}{A}~,
%\label{eq:}
%
\end{align}
up to contact terms which are regularization-dependent.
\item When $\nu=0$,
\begin{align}
\phi \sim A u \ln u (1+\cdots)+ B u(1+\cdots)~,
%\quad (u\to0)~.
%\label{eq:}
%
\end{align}
The Green's function is given by
\begin{align}
\GR = \frac{B}{A}~,
%\label{eq:}
%
\end{align}
up to contact terms.
\end{itemize}
The log term gives logarithmic divergences in the Green's functions, but they can be removed with a boundary counterterm added in the bulk action. However, this necessarily breaks conformal invariance. When $\nu>0$, the relevant counterterm is 
%\footnote{cf. Erdmenger textbook, page 207}
\begin{align}
S_\text{CT} &= -\frac{c_{2\nu}}{2} \int_{\del\calM}d^{p+1}x\, \sqrt{-\gamma} \phi\Box_\gamma^\nu \phi \times \ln\epsilon~, 
%-\frac{c_{2\nu}}{2} &= \frac{1}{2^{2\nu-1}\Gamma(\nu)^2}
%\label{eq:}
%
\end{align}
where $\gamma_{\mu\nu}$ is the $(p+1)$-dimensional boundary metric, $\Box_\gamma$ is the Laplacian made of $\gamma_{\mu\nu}$, and $\epsilon$ is the cutoff. 
When $\nu=0$, the relevant counterterm is 
%\footnote{cf. Skenderis, 0112199 Sec.~5.1.1 and Karch, 0512125}
\begin{align}
S_\text{CT} = -\frac{1}{2} \int_{\del\calM}d^{p+1}x\, \sqrt{-\gamma} \left( \Deltap+\frac{1}{\ln\epsilon} \right) \phi^2~.
%\label{eq:}
%
\end{align}

First, consider the pure AdS$_3$. When $\nu=\mathbb{Z}^+$, the Hankel function is given by 
\begin{subequations}
%\label{eq:}
\begin{align}
H_\nu^{(1,2)} :=& J_\nu \pm iY_\nu 
\label{eq:hankel_integer} \\
\sim & 
\pm \frac{i}{\pi} \biggl[ -\Gamma(\nu) \left(\frac{z}{2}\right)^{-\nu} 
+\cdots
\nonumber \\
&+ \frac{1}{\Gamma(1+\nu)} \left(\frac{z}{2}\right)^\nu 
 \biggl\{ 2 \ln\frac{z}{2} \mp i\pi
 \nonumber \\
& - \psi(1) - \psi(\nu+1)] \biggr\} 
+\cdots \Biggr]
%H_\nu^{(1)} &:= J_\nu + iY_\nu 
%\label{eq:hankel_integer} \\
%\sim & 
%- \frac{i}{\pi}\Gamma(\nu) \left(\frac{z}{2}\right)^{-\nu} 
%+\cdots
%\nonumber \\
%&+ \frac{i}{\pi} \frac{1}{\Gamma(1+\nu)} \left(\frac{z}{2}\right)^\nu 
% \biggl\{ 2 \ln\frac{z}{2} 
% \nonumber \\
%&-i\pi - \psi(1) - \psi(\nu+1)] \biggr\} 
%+\cdots
%\label{eq:}
%
\end{align}
\end{subequations}
instead of \eq{hankel_noninteger}. 
Then, the Green's function is given by%
\footnote{One can get the result from the $\nu\neq\mathbb{Z}$ Green's function as well. Set $\nu=n+\epsilon$ and take the $\epsilon\to0$ limit.}
\begin{align}
\GR =-2\nu \frac{B}{A} = \frac{2}{\Gamma(\nu)^2} \left(\frac{-k^2}{4}\right)^\nu \ln k^2~.
%\label{eq:}
%
\end{align}
When $\nu=0$, 
\begin{subequations}
%\label{eq:}
\begin{align}
H_0^{(1,2)} &\sim \pm \frac{2i}{\pi}\left( \ln z\mp \frac{i\pi}{2}+\gamma-\ln2 \right)+\cdots~,\\
%H_0^{(1)} &\sim 1+\frac{2i}{\pi}(\ln z+\gamma-\ln2)+\cdots~,\\
\GR &= \frac{B}{A}= \frac{1}{2} \ln k^2~.
%\label{eq:}
%
\end{align}
\end{subequations}
% v1.1
Again these results should be understood under the convention \eqref{eq:k^2_rule}.

Now, consider the BTZ black holes. Let us look at how the analysis in \appen{noninteger_Green} fails and look at how the log behavior appears. For both nonextreme and extreme cases, we first construct the incoming-wave solutions. We then expand them asymptotically $z=0$. To do so, we make use of transformations \eqref{eq:Gauss_transf} and \eqref{eq:Whittaker_transf}. However, some care is necessary for integer $\nu$. 

For the nonextreme case, the incoming-wave solution is
\begin{align}
{}_2F_1(a,b,a+b+\nu; 1-z)~.
%\label{eq:}
%
\end{align}
The left-hand side of \eq{Gauss_transf} is well-defined when $\nu=n=\mathbb{Z}$. However, the right-hand side is problematic: the first term contains the factor $\Gamma(-\nu)$ and the second term is proportional to ${}_2F_1(a,b,1-\nu;z)$ which is ill-defined. 

Similarly, for the extreme case, the incoming-wave solution is
\begin{align}
W_{\kappa, \nu/2}(\zeta)~.
%\label{eq:}
%
\end{align}
The left-hand side of \eq{Whittaker_transf} is well-defined when $\nu=n=\mathbb{Z}$, but the right-hand side is problematic: the first term contains the factor $\Gamma(-\nu)$ and the second term is proportional to $W_{\kappa, -n/2}$ which is ill-defined. 

In these cases, set $\nu=n+\epsilon$, substitute power-series expansions of special functions into the transformations, and take the $\epsilon\to0$ limit. The results contain log terms as expected. The results can be found in standard handbooks (\eg, Refs~\cite{nist,abramowitz}), but we collect them in \appen{formula} for reader's convenience.
%In order to obtain Green's functions, we need to extract falloffs $A$ and $B$. When $\nu=n=\mathbb{Z}$, the incoming solution for the nonextreme case is
%\begin{align}
%
%{}_2F_1(a,b,a+b+n; 1-z)~,
%\label{eq:}
%
%\end{align}
%and we need $O(z^n)$ terms. It is useful to use the expressions of the special function in \appen{formula}. 
We use these formulae to extract falloffs $A$ and $B$. 

%Similarly, for the extreme case, the incoming solution is
%\begin{align}
%
%W_{\kappa, n/2}(\zeta)~,
%\label{eq:}
%
%\end{align}
%and we need $O(\zeta^{(1+n)/2})$ terms. Again use the expressions in \appen{formula}. 

Two more remarks: 
\begin{itemize}
\item
The fast falloffs typically contain terms such as $\psi(1)+\psi(n+1)$. 

\item
In the ansatz \eqref{eq:ansatz_stationary}, there is a factor $(1-z)^{-i\lambda}$, and this contributes at $O(z^n)$ because it can be expanded as
\begin{align}
(1-z)^{-i\lambda} = \sum_{j=0}^\infty \frac{ \Gamma(-i\lambda+1) }{ j! \Gamma(-i\lambda-j+1) } (-z)^j~.
%\label{eq:}
%
\end{align}
\end{itemize}
However, these terms gives only contact terms and are ignored.
Anyway, there is no pole-skipping from these contributions.

%%------------------
\subsection{Some useful expressions}\label{sec:formula}
%%------------------

In order to extract falloffs $A,B$, the following expressions of special functions are useful. Below $(a)_j := \tfrac{\Gamma(a+j)}{\Gamma(a)}$ is the Pochhammer symbol and $\psi(z)$ is the digamma function.

%\newpage

%Eq.~(15.8.10) of Ref.~\cite{nist}: 
When $\cn$ is a nonnegative integer, 
\begin{subequations}
%\label{eq:}
\begin{align}
\lefteqn { {}_2F_1(a,b,a+b+\cn; z) =} 
\nonumber \\
& \frac{ (\cn-1)!\Gamma(a+b+\cn) }{ \Gamma(a+\cn)\Gamma(b+\cn) } \sum_{j=0}^{\cn-1} \frac{ (a)_j (b)_j} { j!(1-\cn)_j }  (1-z)^j 
\nonumber \\
& + (z-1)^\cn  \frac{ \Gamma(a+b+\cn) }{ \Gamma(a)\Gamma(b) } 
\sum_{j=0}^{\infty} \frac{ (a+\cn)_j (b+\cn)_j} { j!(j+\cn)! }  (1-z)^j 
\nonumber \\
& \times
[ -\ln(1-z)+\psi(j+1)+\psi(j+\cn+1)
\nonumber \\
&-\psi(a+j+\cn)-\psi(b+j+\cn) ]
\\
\sim 
&\frac{ (\cn-1)! \Gamma(a+b+\cn) }{ \Gamma(a+\cn)\Gamma(b+\cn) } 
+ \cdots 
\nonumber \\
% corrected
& + (z-1)^\cn \frac{ \Gamma(a+b+\cn) }{ \Gamma(a)\Gamma(b)\cn! } 
\nonumber \\
& \times
[ -\ln(1-z)+\psi(1)+\psi(\cn+1)
\nonumber \\
&-\psi(a+\cn)-\psi(b+\cn) ]+\cdots~.
%\label{eq:}
%
\end{align}
\end{subequations}
In the last line, we write only the terms which contribute to the falloffs $A$ and $B$. 
Also, for $\cn=0$, ignore the first sum. So, for $\cn=0$, 
\begin{align}
\lefteqn{ {}_2F_1(a,b,a+b; z) \sim} \\
 & -\frac{ \Gamma(a+b) }{ \Gamma(a)\Gamma(b) } 
[ \ln(1-z)-2\psi(1)+\psi(a)+\psi(b) ] +\cdots~.
\nonumber 
%\label{eq:}
%
\end{align}
%Eq.~(13.14.8), (13.14.9) of Ref.~\cite{nist}: 
%\newpage
When $n$ is an integer and $\kappa-\half\cn-\half\neq 0,1,\ldots,$
\begin{subequations}
\label{eq:Whittaker_integer}
\begin{align}
\lefteqn{ W_{\kappa,\pm\cn/2}(\zeta) =
\frac{ (-)^\cn e^{-\zeta/2} }{ \cn!\Gamma(\half-\half \cn -\kappa)} \zeta^{(1+\cn)/2}
} 
\nonumber \\
& 
\times \left[ \sum_{j=1}^\cn \frac{ \cn!(j-1)!} { (\cn-j)!(\kappa+\half-\half \cn)_j } \zeta^{-j} \right.
\nonumber \\
&- \sum_{j=0}^\infty \frac{ (\half \cn +\half-\kappa)_j }{ (\cn+1)_j j! } \zeta^{j} 
 \\
& \times 
\{\ln \zeta +\psi(\half \cn \!+\! \half \!-\! \kappa \!+\! j )-\psi(j \!+\! 1)-\psi(\cn \!+\! j \!+\! 1) \} \Biggr]
\nonumber \\
\sim & \cdots \Biggl[ (-)^\cn \cn \Gamma(\cn)^2 
\frac{ \Gamma(\half +\half\cn - \kappa) }{ \Gamma(\half - \half\cn - \kappa) } 
\zeta^{(1-\cn)/2} 
+\cdots 
\nonumber \\
& - \zeta^{(1+\cn)/2} 
 \\
&\times \{\ln \zeta+\psi(\half \cn+\half - \kappa)-\psi(1)-\psi(\cn+1) \}  +\cdots \Biggr]~.
\nonumber 
%\label{eq:}
%
\end{align}
\end{subequations}
Again, for $\cn=0$, ignoring the first sum, one gets
\begin{align}
W_{\kappa,0}(\zeta) \sim & 
-\frac{ e^{-\zeta/2} }{ \Gamma(\half-\kappa)} \zeta^{1/2}
%\nonumber \\
%& \times
\{\ln \zeta+\psi(\half-\kappa)-2\psi(1) \}~.
%\label{eq:}
%
\end{align}

%\footnotesize

\end{document}